\documentclass[a4paper,11pt]{article}
\pdfoutput=1

\usepackage{jheppub}
\usepackage[english]{babel}
\usepackage{amsmath, amssymb}
\usepackage{graphicx}
\usepackage{color}
\usepackage{dsfont}
\usepackage{subfigure}

\graphicspath{{./figs/}}

\newcommand{\AddrAHEP}{
  AHEP Group, Institut de F\'{\i}sica Corpuscular --
  C.S.I.C./Universitat de Val{\`e}ncia \\
  Edificio Institutos de Paterna, Apt 22085, E--46071 Valencia, Spain}

\def\SM{$\mathrm{SU(3)_c \otimes SU(2)_L \otimes U(1)_Y}$ }
\newcommand{\sm}{{Standard Model }}
\begin{document}

\title{WIMP dark matter as radiative neutrino mass messenger}

\author[a]{M.~Hirsch}
\author[a]{R.~A.~Lineros}
\author[b]{S.~Morisi}
\author[a]{J.~Palacio}
\author[c]{N.~Rojas}
\author[a]{J.~W.~F.~Valle}
\affiliation[a]{\AddrAHEP}
\affiliation[b]{Institut f{\"u}r Theoretische Physik und Astrophysik,
  Universit{\"a}t W{\"u}rzburg, \\97074 W{\"u}rzburg, Germany.}
\affiliation[c]{Pontificia Universidad Cat\'{o}lica de Chile, Facultad
  de F\'{\i}sica. Av. Vicu\~na Mackenna 4860. Macul. Santiago de
  Chile, Chile.}

\abstract{ The minimal seesaw extension of the Standard \SM Model
  requires two electroweak singlet fermions in order to accommodate
  the neutrino oscillation parameters at tree level.
  Here we consider a next to minimal extension where light neutrino
  masses are generated radiatively by two electroweak fermions: one
  singlet and one triplet under SU(2)$_{\rm L}$.
  These should be odd under a parity symmetry and their mixing gives
  rise to a stable weakly interactive massive particle (WIMP) dark
  matter candidate. For mass in the GeV--TeV range, it reproduces the
  correct relic density, and provides an observable signal in nuclear
  recoil direct detection experiments.
  The fermion triplet component of the dark matter has gauge
  interactions, making it also detectable at present and near future
  collider experiments.  }

\arxivnumber{1307.8134}
\dedicated{IFIC/13-53}

\maketitle

\section{Introduction}

Despite the successful discovery of the Higgs boson, so far the Large
Hadron Collider (LHC) has not discovered any new physics, so neutrino
physics remains, together with dark matter, as the main motivation to
go beyond the Standard Model (SM).
Neutrino oscillation experiments indicate two different neutrino mass
squared differences~\cite{Schwetz:2011zk,Tortola:2012te}. As a result
at least two of the three active neutrino must be massive, though the
oscillation interpretation is compatible with one of the neutrinos
being massless. 
In the \sm neutrinos have no mass at the renormalizable level. However
they can get a Majorana mass by means of the dimension-5 Weinberg
operator,
\begin{equation}
	\label{eq:wo}
	\frac{c}{\Lambda} \, LH\,LH\,,
\end{equation}
where $\Lambda$ is an effective scale, $c$ a dimensionless coefficient
and $L$ and $H$ denote the lepton and Higgs isodoublets, respecively.
This operator should be understood as encoding new physics associated
to heavy ``messenger'' states whose fundamental renormalizable
interactions should be prescribed.
The smallness of neutrino masses compared to the other fermion masses,
suggests that the messenger scale $\Lambda$ must is much higher than
the electroweak scale if the coefficient $c$ in equation~\ref{eq:wo}
is of $\mathcal{O}(1)$.  For example, the scale $\Lambda$ should be
close to the Grand Unification scale if $c$ is generated at tree
level.
One popular mechanism to generate the dimension-5 operator is the
so--called {\it seesaw mechanism}.
Its most general \SM realization is the so called ``1-2-3'' seesaw
scheme~\cite{Schechter:1981cv} with singlet, doublet and triplet
scalar $SU(2)_L$ fields with vevs respectively $v_1$, $v_2$ and $v_3$.
Assuming $m$ extra singlet fermions (right-handed neutrinos), the
``1-2-3'' scheme is described by the $(3+m)\times (3+m)$ matrix 
\begin{equation}
M^\nu = \left(
\begin{array}{cc}
Y_3 v_3 & Y_2 v_2\\
Y_2^T v_2& Y_1 v_1
\end{array}
\right).
\end{equation}
The vevs obey the seesaw relation 
\begin{equation}
v_3 v_1\sim v_2^2\qquad \mbox{with}\qquad v_1\gg v_2 \gg v_3\,,
\end{equation}
giving two contributions to the light neutrino masses $Y_3 v_3 +
v_2^2/v_1\,Y_2Y_1^{-1}Y_2^T$, called respectively type-II and type-I
seesaw.
Assuming $Y_3=0$, namely no Higgs triplet~\footnote{Note that in pure
  type-II seesaw, only one extra scalar field is required, in contrast
  with type-I, where at least two fermion singlets must be assumed.},
the light neutrino masses arise only from the type-I seesaw
contribution.
In this case it is well known that in order to accommodate the
neutrino oscillation parameters, at least two right-handed neutrinos
are required, namely $m\ge 2$. We call the case $m=2$ minimal.
Note that in this case one neutrino mass is zero and so the absolute
neutrino mass scale is fixed.
Typically the next to minimal case is to assume three sequential
right-handed neutrinos, that is $m=3$. 
An alternative seesaw mechanism is the so called type-III in which the
heavy the ``right-handed'' neutrino ``messenger'' states are replaced
by SU(2)$_{\rm L}$ triplet fermions~\cite{Foot:1988aq}. As for the
type-I seesaw case, one must assume at least two fermion triplets (if
only fermion triplets are present) in order to accommodate current
neutrino oscillation data.

There is an interesting way to induce the dimension-5 operator by
mimicking the seesaw mechanism at the radiative level.  This requires
the fermion messengers to be odd under an ad-hoc symmetry $Z_2$ in
order to accommodate a stable dark matter (DM) candidate. In this case
one can have ``scotogenic''~\cite{Ma:2006km} neutrino masses, induced
by dark matter exchange. This trick can be realized either in type-I
or type-III seesaw schemes~\cite{Ma:2006km,Ma:2008cu}.
To induce Yukawa couplings between the extra fermions and the \sm
leptons, one must include additional scalar doublets, odd under the
assumed $Z_2$ symmetry, and without vacuum expectation value.
In order to complete the saga in this paper we propose a hybrid
scotogenic construction which consists in having just one singlet
fermion ($m=1$) but adding one triplet fermion as well.
\begin{figure}[tb]
\begin{center}
\includegraphics[width=0.5\textwidth]{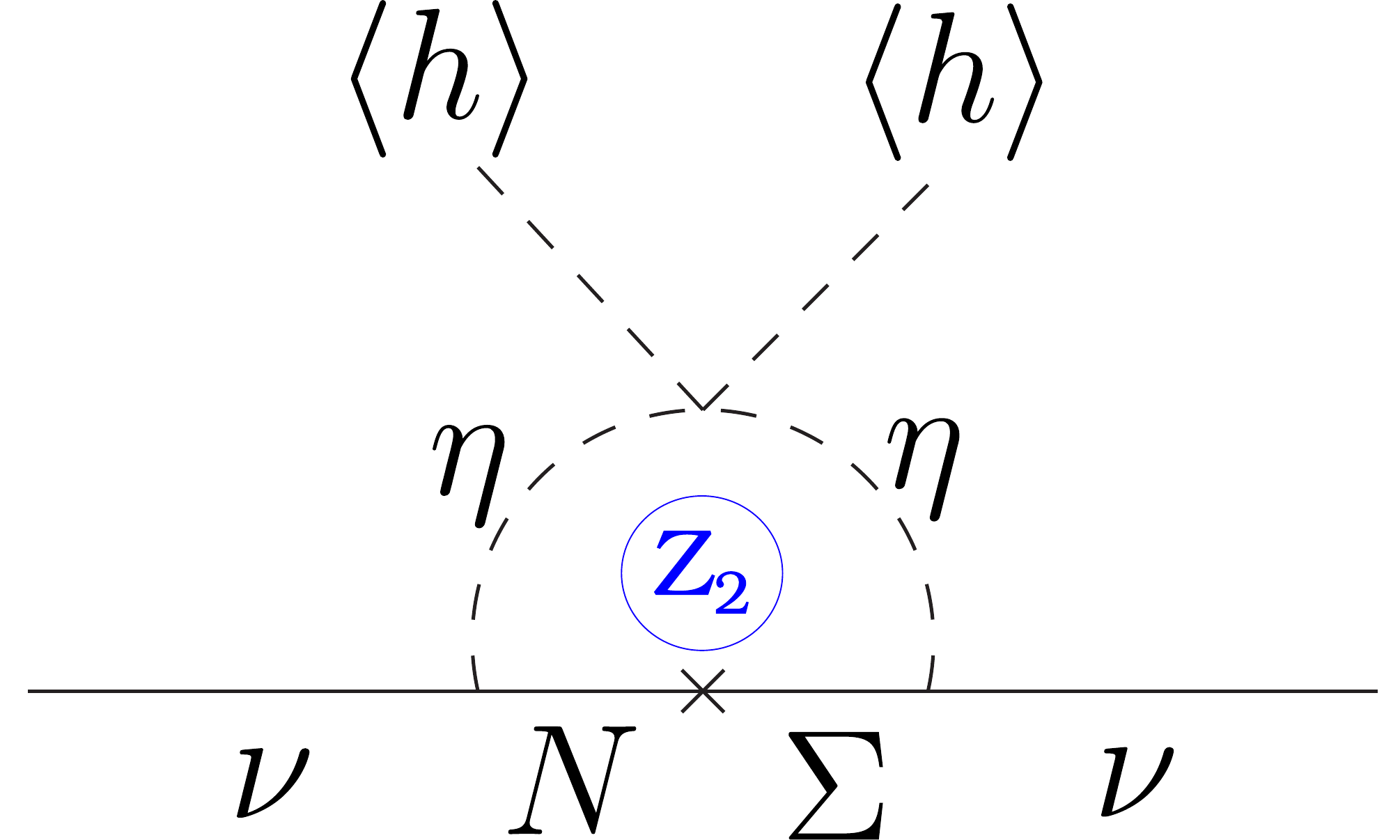}
\caption{One loop realization for the Weinberg operator.}
\label{fig:radneutmass}
\end{center}
\end{figure}
This also gives rise to light neutrino masses, calculable at the one
loop level, as illustrated in
figure~\ref{fig:radneutmass}~\footnote{Note the scalar contributions
  come from the scalar and pseudoscalar pieces of the field $\eta$.}.
However, due to triplet--singlet mixing, the lightest combimation of
the neutral component of the fermion triplet and the singlet will be
stable and can play the role of WIMP dark matter. We show that it
provides a phenomenologically interesting alternative to all previous
``scotogenic'' proposals since here the dark matter can have sizeable
gauge interactions. As a result, in addition to direct and indirect
detection signatures, it can also be kinematically accessible to
searches at
present colliders such as the LHC.\\

Existing collider searches at LEP~\cite{Ellis:1988zy,L3:2001PhLB} and
LHC~\cite{CMS:2012PhLB}, set a nominal lower bound of $\sim$ 100 GeV
for the masses of new charged particles.
However, coannihilations present in the early universe, between the
neutral and charged components, set the dark matter mass to be of the
order of~\cite{Ma:2008cu} 
\begin{equation}
  \label{eq:33DM}
M_{\rm DM} \simeq 2.7~{\rm TeV}  
\end{equation}
in order to explain the observed abundance~\cite{planck:2013}:
\begin{equation}
	\label{eq:omegaDM}
	\Omega_{\rm DM} h^2 = 0.1196 \pm 0.0031 \, .
\end{equation}

Radiative neutrino masses generated by at least two generations of
fermion singlets or triplets have been studied in
Ref.~\cite{Kubo:2006yx}.
Here we focus on the radiative neutrino mass generation with one
singlet and one triplet fermion which has interesting phenomenological
consequences compared to the cases aforementioned cases.
In our scenario, the dark matter candidate can indeed be observed not
only in indirect but can also be kinematically accessible to current
collider searches, and need not obey Eq.~(\ref{eq:33DM}). Moreover, we
will show that, in contrast to the proposed schemes in
Refs.~\cite{Ma:2006km,Ma:2008cu} in our framework amplitudes leading
naturally to direct detection processes appear at the tree level,
thanks to singlet-triplet mixing effects.


The rest of this paper is organized as follows: in
section~\ref{sec:model} we introduce the new fields and interactions
present in the model, making emphasis upon the mixing matrices and the
radiative neutrino mass generation mechanism.
Section~\ref{sec:DMsect} is devoted to numerical results on the
phenomenology of dark matter in this model.
An interesting feature of the model is the wide range of possible dark
matter masses, ranging from 1~GeV to a few TeV. We also briefly
discuss some the implications for LHC physics.
In Section~\ref{sec:concl} we give our conclusions.


\section{The model}
\label{sec:model}


Our model combines the ingredients employed in the models
proposed in~\cite{Ma:2006km,Ma:2008cu} in such a way that it has a richer
phenomenology than either~\cite{Ma:2006km} or~\cite{Ma:2008cu}.

\subsection{The Model and the Particle Content}

The new fields with respect to the \sm include one Majorana fermion
triplet $\Sigma$ and a Majorana fermion singlet $N$ both with zero
hypercharge and both odd under an ad-hoc symmetry $Z_2$.  We also
include a scalar doublet $\eta$ with same quantum numbers as the Higgs
doublet, but odd under $Z_2$.  In addition, we require that $\eta$ not
to acquire a vev.
As a result, neutrino masses are not generated at tree level by a
type-I/III seesaw mechanism. Instead they are one-loop calculable,
from diagrams in  Fig.~\ref{fig:radneutmass}.
Furthermore, this symmetry forbids the decays of the lightest $Z_2$
odd particle into \sm particles, which is a mixture of the neutral
component of $\Sigma$ and $N$. As a result this becomes a viable dark
matter candidate. Note also that our proposed model does not modify
quark dynamics, since neither of the new fields couples to quarks.\\

The fermion triplet, can be expanded as follows ($\sigma_i$ are the
Pauli matrices):
\begin{eqnarray}
\Sigma &=& \Sigma_1 \sigma_1 + \Sigma_2 \sigma_2 + \Sigma_3 \sigma_3 \, =\, \left(
	\begin{array}{cc}
	\Sigma_0 & \sqrt{2}\Sigma^{+} \\ 
	\sqrt{2}\Sigma^{-} & -\Sigma_0 \\
	\end{array}
\right) \, ,
\end{eqnarray}
where
\begin{eqnarray}
	\Sigma^+ &=& \frac{1}{\sqrt{2}}\left(\Sigma_1+i\Sigma_2\right)\, , \\
	\Sigma^-&= &\frac{1}{\sqrt{2}}\left(\Sigma_1-i\Sigma_2\right)\,,\\
	\Sigma^0&= & \Sigma_3 \, .
\end{eqnarray}

The $Z_2$ is exactly conserved in the Lagrangian, moreover, it allows
interactions between dark matter and leptons, in fact, this is the origin of
radiative neutrino masses.
The Yukawa couplings between the triplet and leptons play an important
role in the dark matter production.
Finally a triplet scalar $\Omega$ is introduced in order to mix the
neutral part of the fermion triplet $\Sigma^0$ and the fermion singlet
$N$.
This triplet scalar field also has zero hypercharge and is even under
the $Z_2$ symmetry, thus, its neutral component can acquire a nonzero
vev.

\begin{table}[tb]
\centering
\begin{tabular}{|c||c|c|c||c|c||c|c|}
\hline
        & \multicolumn{3}{|c||}{Standard Model} &  \multicolumn{2}{|c||}{Fermions}  & \multicolumn{2}{|c|}{Scalars}  \\
        \cline{2-8}
        &  $L$  &  $e$  & $\phi$  & $\Sigma$ &  N   & $\eta$ & $\Omega$ \\
\hline                                                                  
$SU(2)_L$ &  2    &  1    &    2    &     3    &  1   &    2   &    3     \\
$Y$     & -1    &  -2    &    1    &     0    &  0   &    1   &    0     \\
$Z_2$   &  $+$  &  $+$  &   $+$   &    $-$   & $-$  &   $-$  &   $+$    \\
\hline
\end{tabular}
\caption{Matter assignment of the model.}
\label{tab:MatterModel}
\end{table}

\subsection{Yukawa Interactions and Fermion Masses}

The most general \SM and Lorentz invariant Lagrangian is given as
\begin{eqnarray}
\mathcal{L} &\supset& - Y_{\alpha \beta}\,\overline{L}_{\alpha}  e_{\beta} \phi- Y_{\Sigma_\alpha} \overline{L}_{\alpha}C\Sigma^{\dagger} \tilde{\eta} - 
\frac{ 1 }{4}M_\Sigma \mbox{Tr} \left[\overline{\Sigma}^{c} \Sigma \right] + \nonumber \\
& & -Y_{\Omega} \mbox{Tr}\left[ \overline{\Sigma} \Omega \right]N - Y_{N_\alpha} \overline{L}_{\alpha}\tilde{\eta} N - \frac{1}{2}M_N \overline{N}^{c}N 
+ h.c.\, , \label{eq:lagrangian}
\label{eq:lagrangian}
\end{eqnarray}
The $C$ symbol stands for the Lorentz charge conjugation
matrix $i\sigma_2$ and $\tilde{\eta} = i\sigma_2 \eta^{*}$.

The Yukawa term $Y_{\alpha \beta}$ is the SM Yukawa interaction for
leptons, taken as diagonal matrix in the flavor basis\footnote{We can
  always go to this basis with a unitary transformation.}.
On the other hand the Yukawa coupling $Y_{\Omega}$ mixes the $\Sigma$
and $N$ fields and when the neutral part of the $\Omega$ field acquire
a vev $v_\Omega$, the dark matter particle can be identified to the
lightest mass eigenstate of the mass matrix,
\begin{eqnarray}
\label{eq:chimass}
M_{\chi} &=& \left(
             \begin{array}{cc}
             M_\Sigma & 2 Y_{\Omega}v_\Omega \\
             2 Y_{\Omega}v_\Omega  & M_N
             \end{array} 
             \right)\, , 
\end{eqnarray}
in the basis $\psi^T = \left( \Sigma_0\, , N \right)$. As a result one
gets the following tree level fermion masses
\begin{eqnarray}
	m_{\chi^\pm} &=& M_{\Sigma} \, ,\\
	m_{\chi^0_1} &=& \frac{1}{2}\left( M_\Sigma + M_N - \sqrt{\displaystyle (M_\Sigma - M_N )^2 + 4 (2 Y_\Omega v_\Omega)^2} \right) \, ,\\
	m_{\chi^0_2} &=& \frac{1}{2}\left( M_\Sigma + M_N + \sqrt{\displaystyle (M_\Sigma - M_N )^2 + 4 (2 Y_\Omega v_\Omega)^2} \right) \, ,\\
	\tan(2 \, \alpha) &=& \frac{4 Y_\Omega v_\Omega}{M_\Sigma - M_N} \, ,
\end{eqnarray}
where $\alpha$ is the mixing angle between $\Sigma_0$ and $N$.
Here $M_\Sigma$ and $M_N$ characterize the Majorana mass terms for the
triplet and the singlet, respectively.  The $M_\Sigma$ term is also
the mass of the charged component of the $\Sigma$ field, this issue is
important because the mass splitting between $\Sigma^{\pm}$ and the
dark matter candidate will play a role in the calculation of its relic
density. As we will see later, the splitting induced by $v_{\Omega}$
allows us to relax the constraints on the dark matter coming from the
existence of $\Sigma^{\pm}$.\\


\subsection{Scalar potential and spectrum}
\label{sec:scalars}

The most general scalar potential, even under $Z_2$, including the
fields $\phi$, $\eta$ and $\Omega$ and allowing for spontaneous
symmetry breaking, may be written as:
\begin{eqnarray}
  V_{\rm scal} &=& -m_{1}^2 \phi^\dagger \phi + m_{2}^2 \eta^\dagger \eta + \frac{\lambda_1}{2} \left( \phi^\dagger \phi \right)^2 + \frac{\lambda_2}{2} \left( \eta^\dagger \eta \right)^2 + \lambda_3 \left( \phi^\dagger \phi \right)\left( \eta^\dagger \eta \right) \nonumber \\ 
  &+& \lambda_4 \left( \phi^\dagger \eta \right)\left( \eta^\dagger \phi \right) + \frac{\lambda_5}{2} \left(\phi^\dagger \eta \right)^2 + h.c. - \frac{M_\Omega^2}{4} Tr \left( \Omega^\dagger \Omega \right) + \left( \mu_1 \phi^\dagger \Omega \phi + h.c. \right) \nonumber \\
  &+& \lambda^{\Omega}_1 \phi^\dagger \phi \, Tr\left( \Omega^\dagger \Omega\right) + \lambda^{\Omega}_2 \left( Tr (\Omega^\dagger \Omega ) \right)^2 + \lambda^{\Omega}_3 Tr ( \left( \Omega^\dagger \Omega \right)^2 ) + \lambda^{\Omega}_4 \left( \phi^\dagger \Omega \right) \left( \Omega^\dagger \phi \right) \nonumber \\
  &+& \left( \mu_2 \eta^\dagger \Omega \eta + h.c.  \right) + \lambda^{\eta}_1 \eta^\dagger \eta \, Tr\left( \Omega^\dagger \Omega\right) + \lambda^{\eta}_4 \left( \eta^\dagger \Omega \right) \left( \Omega^\dagger \eta \right)\, , \label{eq:scpot}
\end{eqnarray}
where the fields $\eta$, $\phi$ and $\Omega$, can be written as
follows:
\begin{eqnarray}
\eta &=& \left(
         \begin{array}{c}
         \eta^{+} \\
         (\eta^{0} + i\eta^{A})/\sqrt{2}
         \end{array}
         \right) \, ,\nonumber \\
\phi &=& \left(
         \begin{array}{c}
         \varphi^{+} \\
         (h_0 + v_h + i\varphi)/\sqrt{2}
         \end{array}
         \right) \, , \nonumber \\
\Omega &=& \left(
           \begin{array}{cc}
           (\Omega_0 + v_\Omega) & \sqrt{2}\,\Omega^{+} \\
           \sqrt{2}\,\Omega^{-} & - (\Omega_0 + v_\Omega)
           \end{array}
           \right)\, ,
\end{eqnarray}
where $v_{h}$ and $v_{\Omega}$ are the vevs of $\phi$ and $\Omega$
fields respectively.
We have three charged fields one of which is absorbed by the $W$
boson, three CP-even physical neutral fields, and two CP-odd neutral
fields one of which is absorbed by the $Z$ boson~\footnote{Remember
  that the neutral part of $\Omega$ field is real, so it does not
  contribute to the CP-odd sector.}.

Let us first consider the charged scalar sector.
The charged Goldstone boson is a linear combination of the
$\varphi^{+}$ and the $\Omega^{+}$, changing the definition for the
$W$ boson mass from that in the \sm: $\displaystyle M_W =
\frac{g}{2}\sqrt{v_h^2 + v_\Omega^2}$.
Note that this places a constraint on the vev of $v_\Omega$ from
electroweak precision tests~\cite{Gunion:1989ci,Gunion:1989we}, one
can expect roughly this vev to be less than 7~GeV, in order to keep
the $\displaystyle M_Z = \frac{\sqrt{g^2 + {g^{\prime}}^2}}{2}v_h$ in
the experimental range, and alter the $M_W$ value inside the
experimental error band.\\

Apart from the $W$ boson, the two charged scalars have mass:
\begin{eqnarray}
M_{\pm}^2 & = &2\mu_1 \left( v_h^2 + v_\Omega^2\right)/v_\Omega \, ,\\
m_{\eta^\pm}^2 &=& m_2^2 + \frac{1}{2} \lambda_3 v_h^2 + 2\mu_2 v_\Omega +\left(2\lambda^{\eta}_1 + \lambda^{\eta}_4\right)v_\Omega^2\, .
\end{eqnarray}
Notice that the nonzero vacuum expectation value $v_{\Omega} \neq 0$
will play an important role in generating the novel phenomenological
effects of interest to us (see below).
Now let us consider the neutral part: the minimization conditions of
the Higgs potential allow vevs for the neutral part of the usual
$\phi$ field as well as for the neutral part of the $\Omega$ field.
The mass matrix for neutral scalar eigenstates in the basis $\Phi^T =
\left( h_0\, , \Omega_0 \right)$ is:
\begin{eqnarray}
\mathcal{M}_{s}^2 &=& \left( 
                      \begin{array}{cc}
                      \lambda_1 v_h^2 + \frac{t_h}{v_h} &  -2\mu_1 v_h + 4 v_h v_\Omega \left( \lambda^{\Omega}_1 + \frac{\lambda^{\Omega}_4}{2}\right) \\
                      -2\mu_1 v_h + 4 v_h v_\Omega \left( \lambda^{\Omega}_1 + \frac{\lambda^{\Omega}_4}{2}\right) & \frac{\mu_1 v_h^2}{v_\Omega} + 16 v_{\Omega}^2 \left( 2\lambda^{\Omega}_2 +\lambda^{\Omega}_3 \right) + \frac{t_\Omega}{v_\Omega}  \\
                      \end{array}
                      \right) \, , \label{eq:evenscalarmass}
\end{eqnarray}
where $t_h$ and $t_{\Omega}$ are the tadpoles for $h_0$ and $\Omega_0$
and are described in Appendix~\ref{app:tad}. The presence of the vev
$v_\Omega$ induces the mixing between $h_0$ and $\Omega_0$. The
corresponding eigenvalues give us the masses of the \sm Higgs doublet
and the second neutral scalar both labelled as $S_i^0$.

On the other hand, the $\eta$ field does not acquire vev, therefore,
the mass eigenvalues of the neutral $\eta^0$, charged $\eta^{\pm}$ and
pseudoscalar $\eta^A$ are decoupled. The spectrum for $\eta^0$ and
$\eta^A$ fields is:
\begin{eqnarray}
	m_{\eta 0}^2 &=& m_{\eta \pm}^2 + \frac{1}{2}\left( \lambda_4 + \lambda_5 \right) v_h^2 - 4 \mu_2 v_\Omega \, ,\label{eta0}\\
	m_{\eta A}^2 &=& m_{\eta \pm}^2 + \frac{1}{2}\left( \lambda_4 - \lambda_5 \right) v_h^2 - 4 \mu_2 v_\Omega\, . \label{etaA}
\end{eqnarray}


\subsection{Radiative Neutrino Masses}

In this model, neutrino masses are generated at one loop. The dark
matter candidate particle acts as a messenger for the masses.
The relevant interactions for radiative neutrino mass generation arise
from from Eqs.~(\ref{eq:lagrangian}) and (\ref{eq:scpot}) and can be
written in terms of the tree level mass eigenstates. Symbolically, one
can rewrite the relevant terms for this purpose as:
\begin{eqnarray}
\begin{array}{ccccc}
   L\, \Sigma\, \eta                  & \longrightarrow &  h_{ij} \nu_i \, {\chi}^0_{j}\, \eta_0 \,                 & , & h_{ij} \nu_i\, {\chi}^0_{j}\, \eta_A    \\
   L\, \eta\, N                       & \longrightarrow &  h_{ij} \nu_i\, {\chi}^0_{j}\, \eta_0 \,                 & , & h_{ij} \nu_{i}\, {\chi}^0_{j}\, \eta_A    \\
   \left(\phi^\dagger \eta \right)^2  & \longrightarrow &  \left[ \left( h + v_h \right)\,\eta_0 \right]^2 \,   & , & \left[ \left( h + v_h \right)\,\eta_A \right]^2 \\
\end{array}
\end{eqnarray}
{}Here the field ${\chi}^0_{j}$ are the mass eigenstate of the matrix
(\ref{eq:chimass}) and $h$ is a $3\times 2$ matrix and is given by
\begin{equation}
h= \left(
\begin{array}{cc}
Y_{1}^\Sigma & Y_{1}^N  \\
Y_{2}^\Sigma & Y_{2}^N  \\
Y_{3}^\Sigma & Y_{3}^N  \\
\end{array}
\right)
\cdot V(\alpha)\, . \label{eq:hdef}
\end{equation}
where $V(\alpha)$ is the $2\times 2 $ orthogonal matrix that
diagonalizes the matrix in equation~(\ref{eq:chimass}).  There are two
contributions to the neutrino masses from the loops in figure
\ref{fig:radneutmass}, where the $\eta_0$ and $\eta_A$ fields are
involved in the loop.
With the above ingredients, from the diagram in
Fig.~\ref{fig:radneutmass} one finds that the neutrino mass matrix is
given by:
\begin{eqnarray}
M_{\alpha\beta}^\nu &=& \sum_{k = 1,2} \frac{h_{\alpha \sigma}h_{\beta \sigma}}{16\pi^2}I_k \left( M_k, m^2_{\eta_0} , m^2_{\eta_A}\right) \, . \label{eq:numass}
\end{eqnarray}
The $I_{k}$ functions correspond essentially to a differences of the
$B_0$ Veltman functions \cite{Passarino:1978jh}, when evaluated at
different scalar masses, note they have mass dimensions. The index $k$
runs over the $\chi^0$ mass eigenvalues, i.e. $\sigma=1,2$.
Note that these masses are independent of the renormalization
scale. In the equation below, each $M_k$ stands for the mass values of
the $\chi^0$ fields.
\begin{eqnarray}
I_k \left( M_k, m^2_{\eta_0} , m^2_{\eta_A}\right) = M_k\frac{m^2_{\eta_0}}{m_{\eta_0}^2-M_k^2}\log \left( \frac{m_{\eta_0}^2}{M_k^2} \right) - M_k\frac{m^2_{\eta_A}}{m_{\eta_A}^2-M_k^2}\log \left( \frac{m_{\eta_A}^2}{M_k^2} \right) \label{eq:Iformula}
\end{eqnarray}

It is useful to rewrite the equation \ref{eq:numass} in a compact way
as follows
\begin{eqnarray}
M^{\nu} &=& hv_h \cdot \left(	
                     \begin{array}{cc}
                     \frac{I_1}{16\pi^2 v_h^2} & 0 \\
                     0   & \frac{I_2}{16\pi^2 v_h^2}
                     \end{array}
                     \right) \cdot h^T v_h \equiv h v_h \cdot \frac{D_I}{v_h^2} \cdot h^T v_h \sim m_D \frac{1}{M_R}m_D^T\label{eq:numass2}
\end{eqnarray}
which is formally equivalent to the standard type-I seesaw relation
with $M_R^{-1}\to D_I/v_h^2$~\cite{Schechter:1980gr}. This is a
diagonal matrix while $h\,v_h$ plays the role of the Dirac mass
matrix, in our case it is a $3\times 2$ matrix.  It is not difficult
to see that we can fit the required neutrino oscillation
parameters~\cite{Schwetz:2011zk,Tortola:2012te}, for example, by means
of the Casas Ibarra parametrization~\cite{Casas:2001sr}.

In order to get an idea about the order of magnitude of the parameters
required for producing the correct neutrino masses, one can consider a
special limit in equation~\ref{eq:numass}.  For example, in cases
where both $\chi^0$ are lighter than the other fields, we have from
\ref{eq:numass}:
\begin{eqnarray}
M_{\alpha\beta}^\nu &=& \sum_{\sigma = 1,2} \frac{h_{\alpha \sigma}h_{\beta \sigma}}{8\pi^2} \frac{\lambda_5 v_h^2}{m_0^2}M_k \, .\label{eq:numass3}
\end{eqnarray}
Here $\lambda_5$ is the $\left( \phi^\dagger \eta \right)^2$ coupling
introduced in equation \ref{eq:scpot}.  The $M_k$ are the masses of
the neutral $Z_2$ fermion fields $\chi$.
The $m_0$ mass term
comes from writing the masses of the $\eta_0$, and $\eta_A$ in the following 
way: $m^2_{\eta_0,\,\eta_A} = m_0 \pm \lambda_5 v_h^2$, see appendix 
\ref{sec:approx} for more details.
In particular we are interested in the magnitude of the Yukawa
couplings $h_{\alpha\beta}$ required in order to have neutrino with
masses of the order of eV.  For masses of $\chi^0$ of order of 10 GeV
and $\eta_{0,A}$ of order of 1000 GeV, and $\lambda$ couplings not too
small, namely of order of $10^{-2}$, one finds that the values for
$h_{\alpha\beta}$ are in the order of the bottom Yukawa coupling $\sim
10^{-2}$. Hence it is not necessary to have a tiny Yukawa for
obtaining the correct neutrino masses.

\section{Fermion Dark Matter}
\label{sec:DMsect}

\begin{figure}[tb]
\centering
\subfigure[]{
		\centering
		\includegraphics[width = 0.35\textwidth]{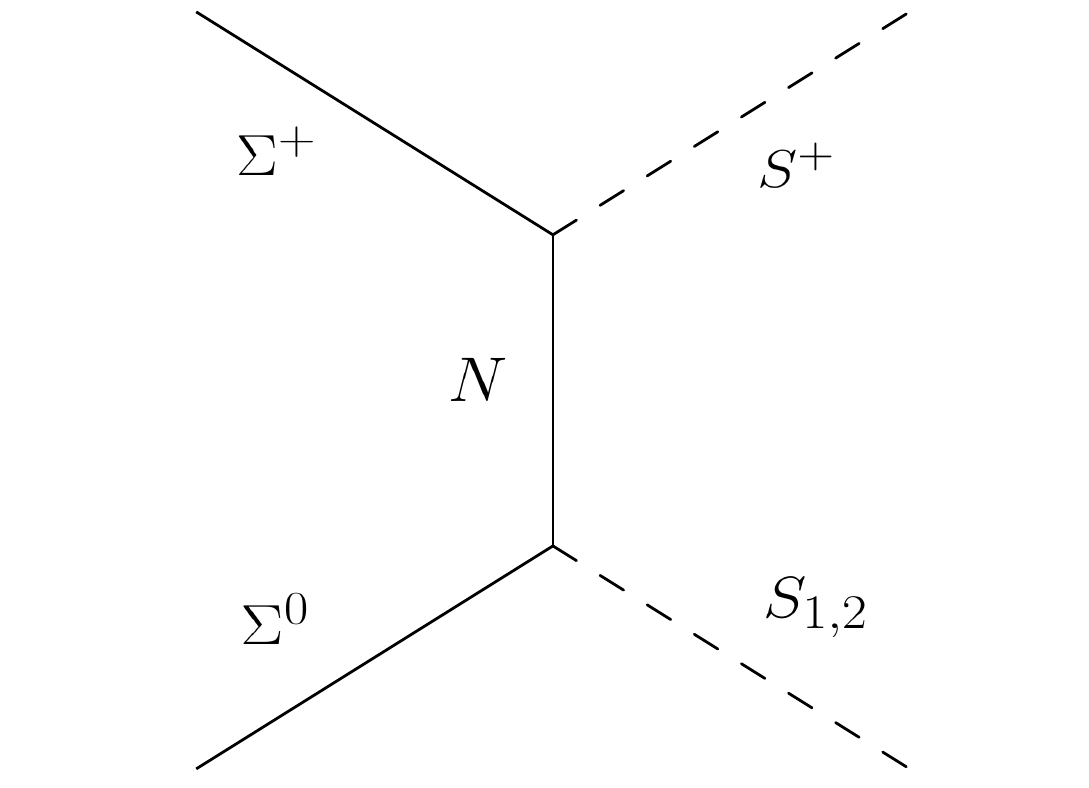}}
\subfigure[]{
		\centering
		\includegraphics[width = 0.35\textwidth]{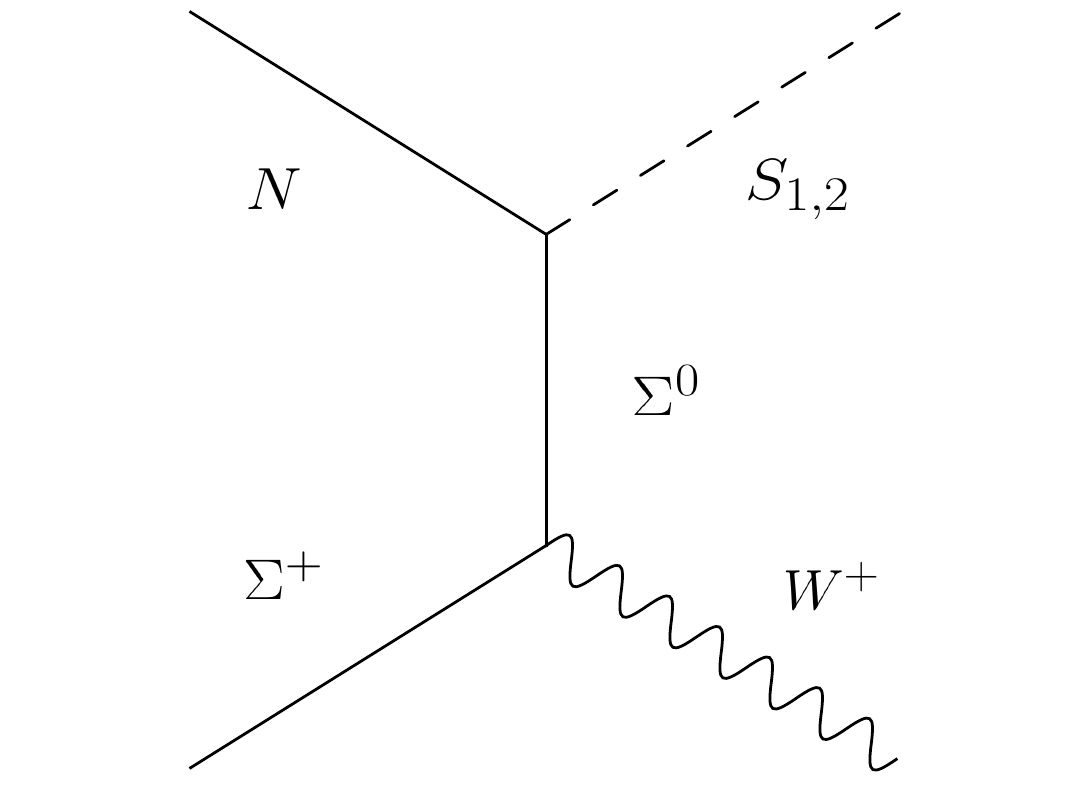}}

\subfigure[]{
		\centering
		\includegraphics[width = 0.35\textwidth]{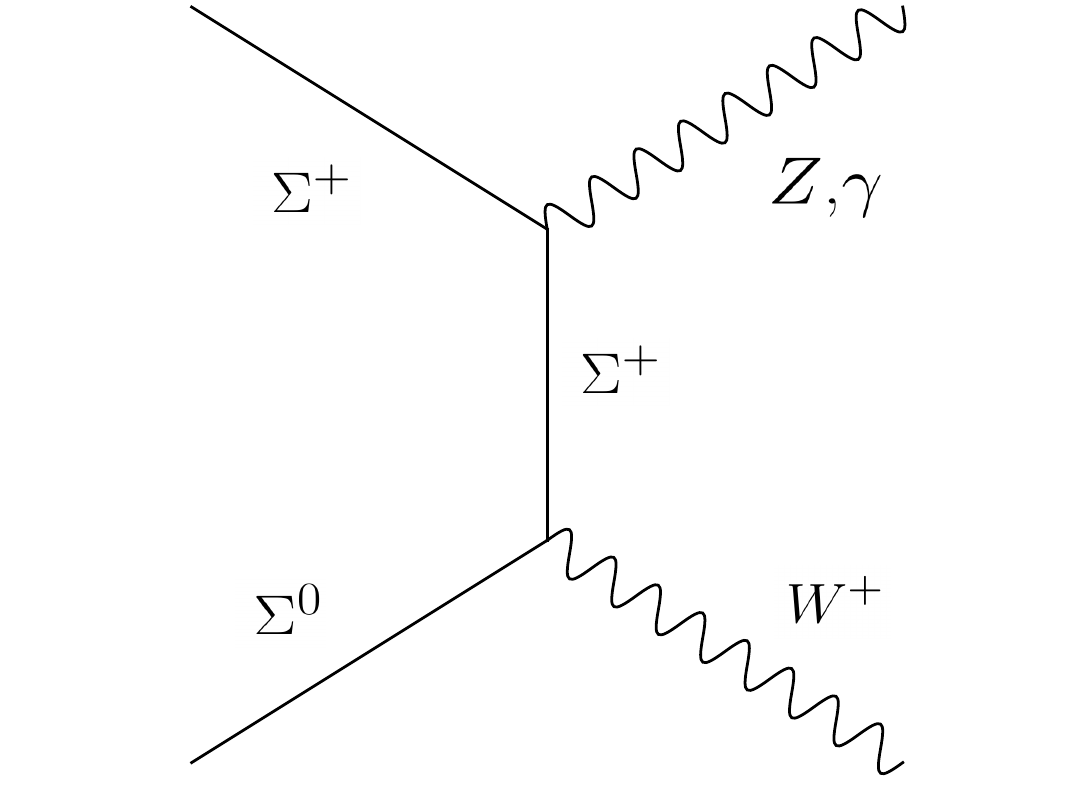}}
\subfigure[]{
		\centering
		\includegraphics[width = 0.35\textwidth]{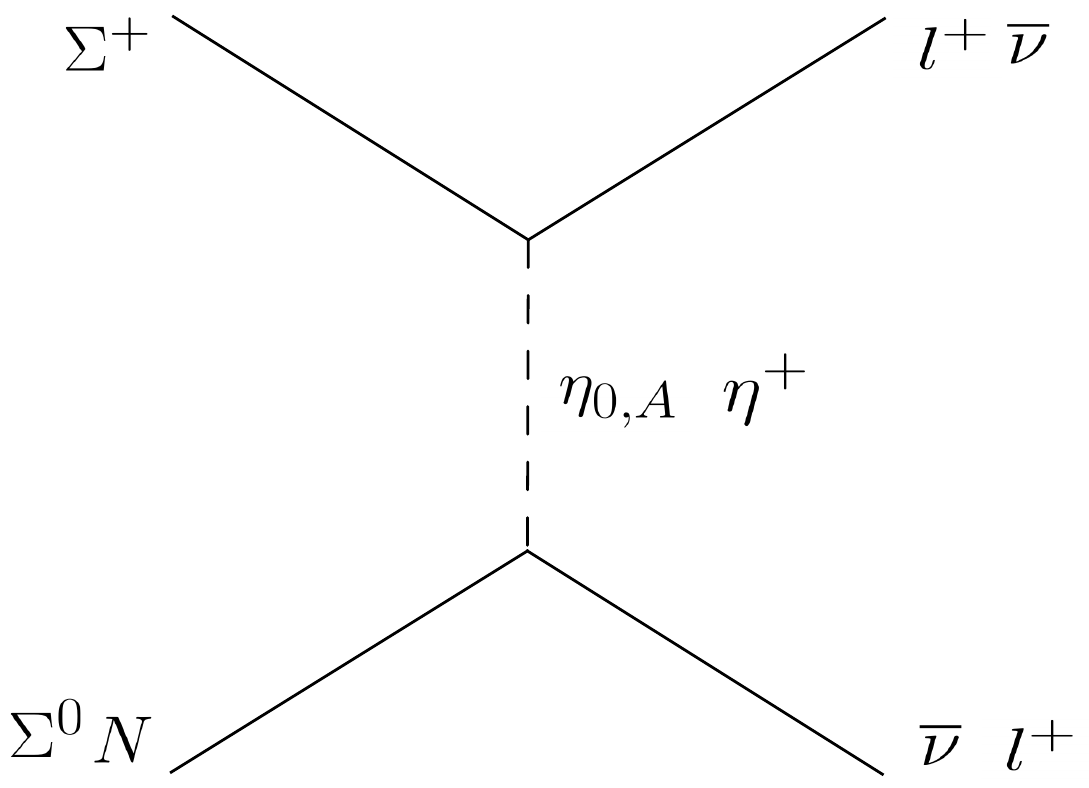}}

\subfigure[]{
		\centering
		\includegraphics[width = 0.35\textwidth]{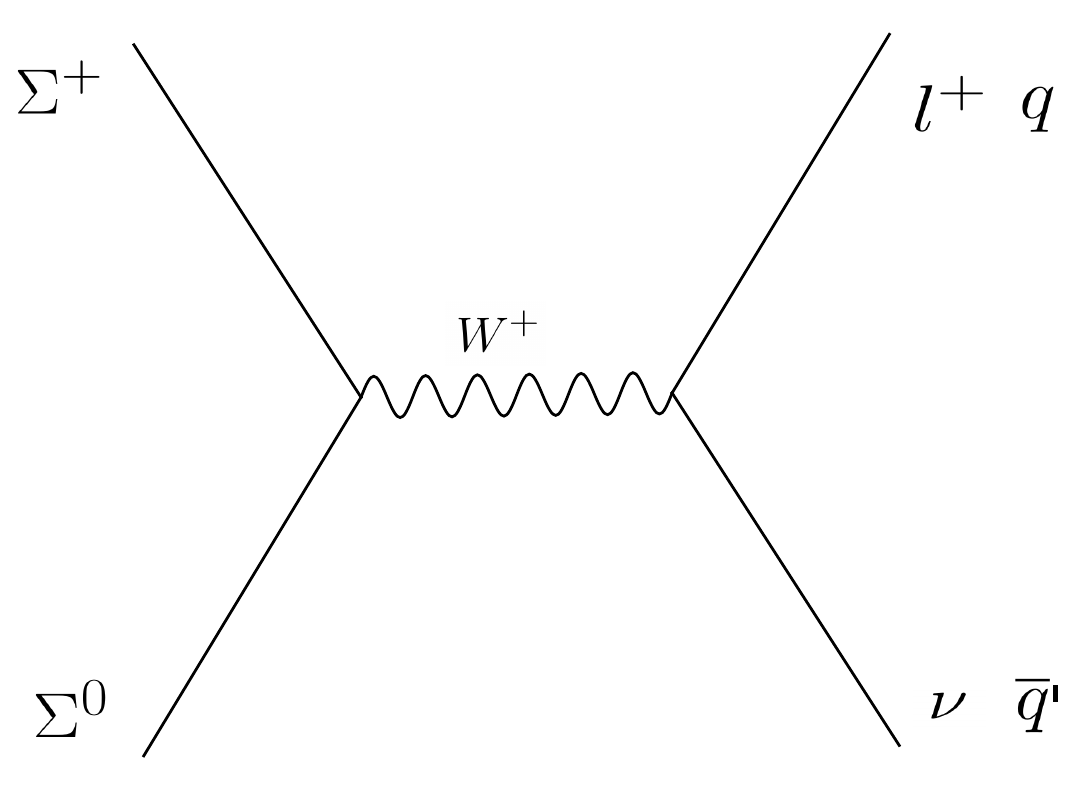}}
\subfigure[]{
		\centering
		\includegraphics[width = 0.35\textwidth]{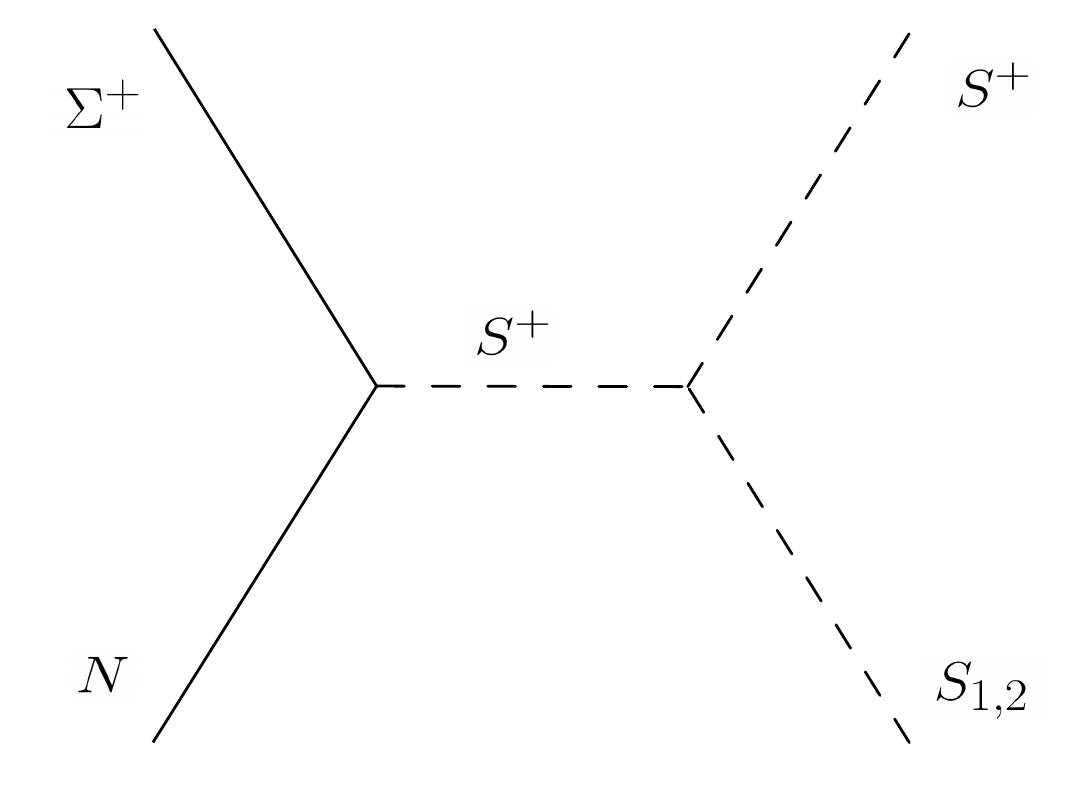}}

\subfigure[]{
		\centering
		\includegraphics[width = 0.35\textwidth]{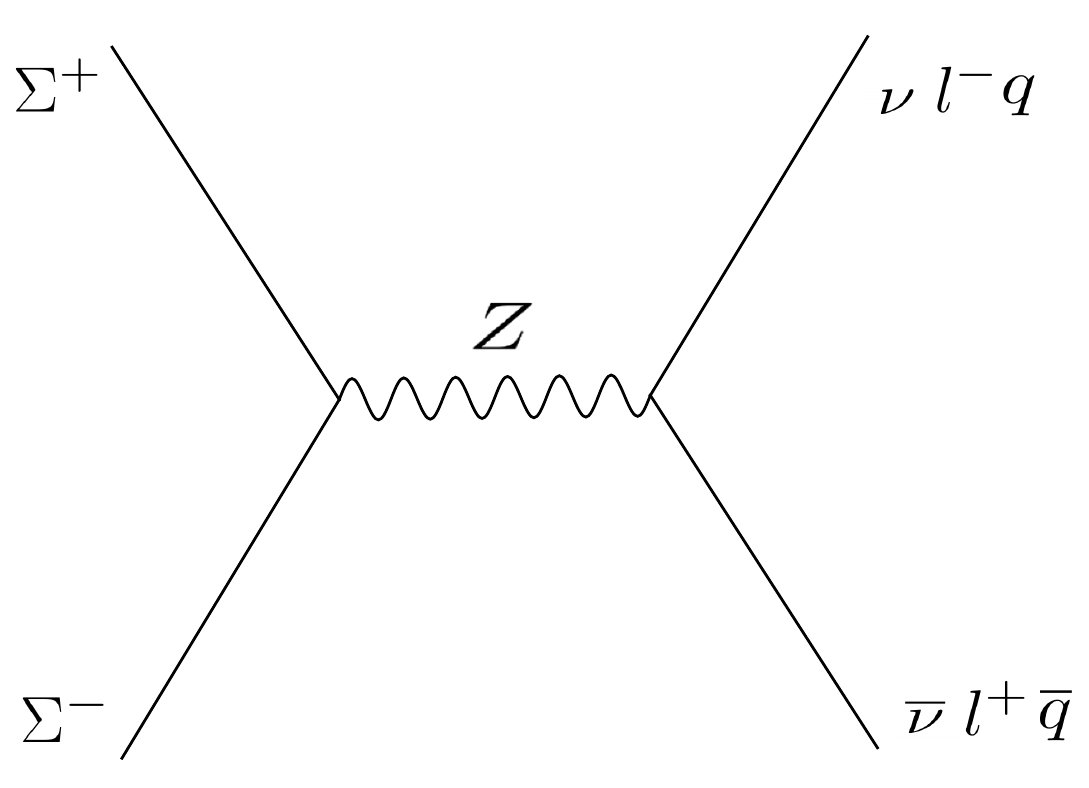}}
\subfigure[]{
		\centering
		\includegraphics[width = 0.35\textwidth]{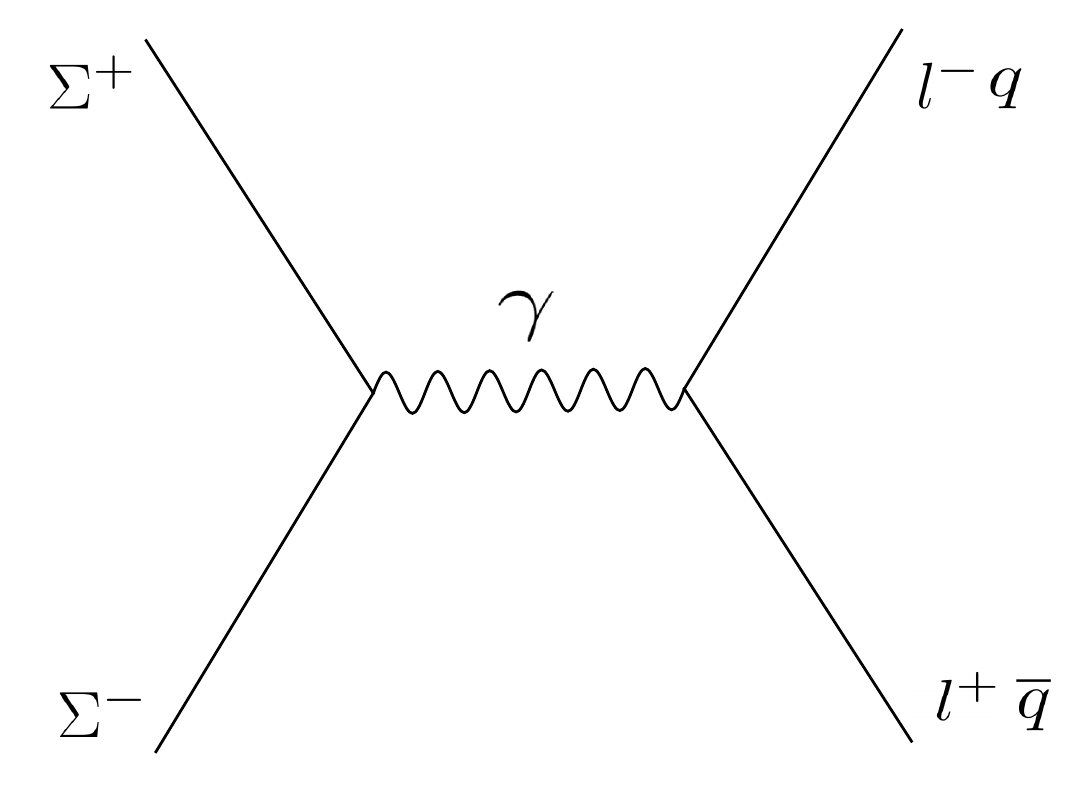}}

	\caption{\label{fig:coann}
		$\Sigma^0$ and $N$ co-annihilation channels.
		Figures (g) and (h) correspond to the processes involved in the $\Sigma^{\pm}$ abundace.}
\end{figure}

\begin{figure}[tb]
\centering
\subfigure[]{
		\centering
		\includegraphics[width = 0.35\textwidth]{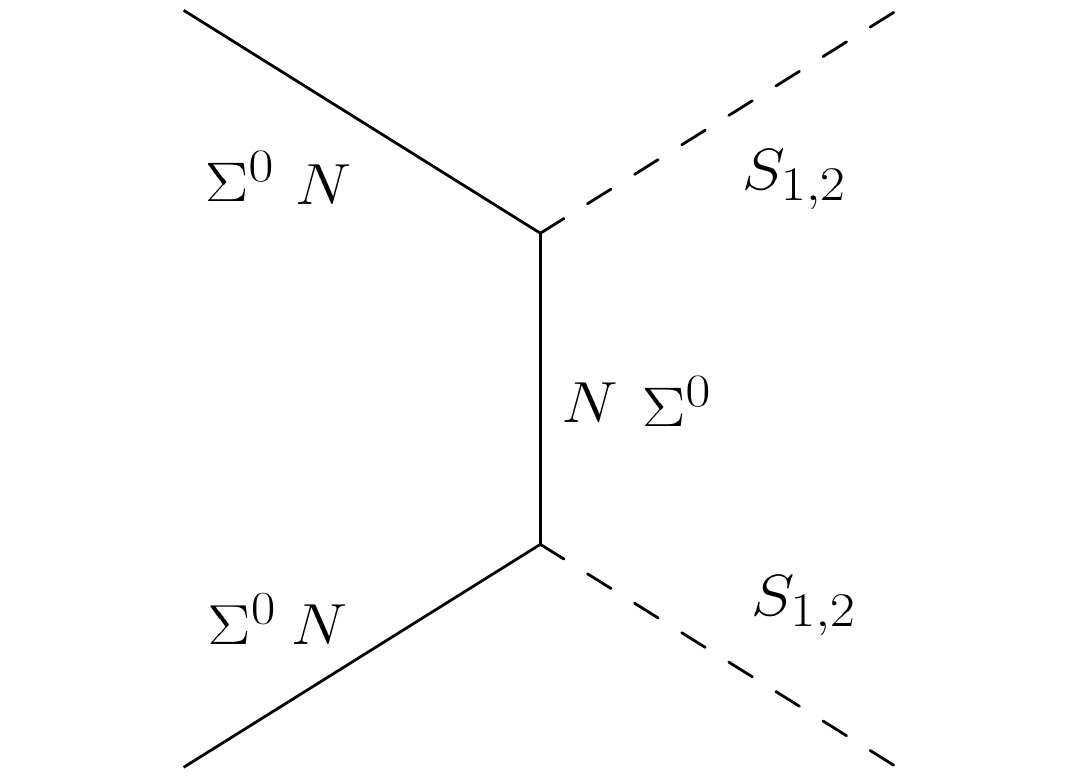}}
\subfigure[]{
		\centering
		\includegraphics[width = 0.35\textwidth]{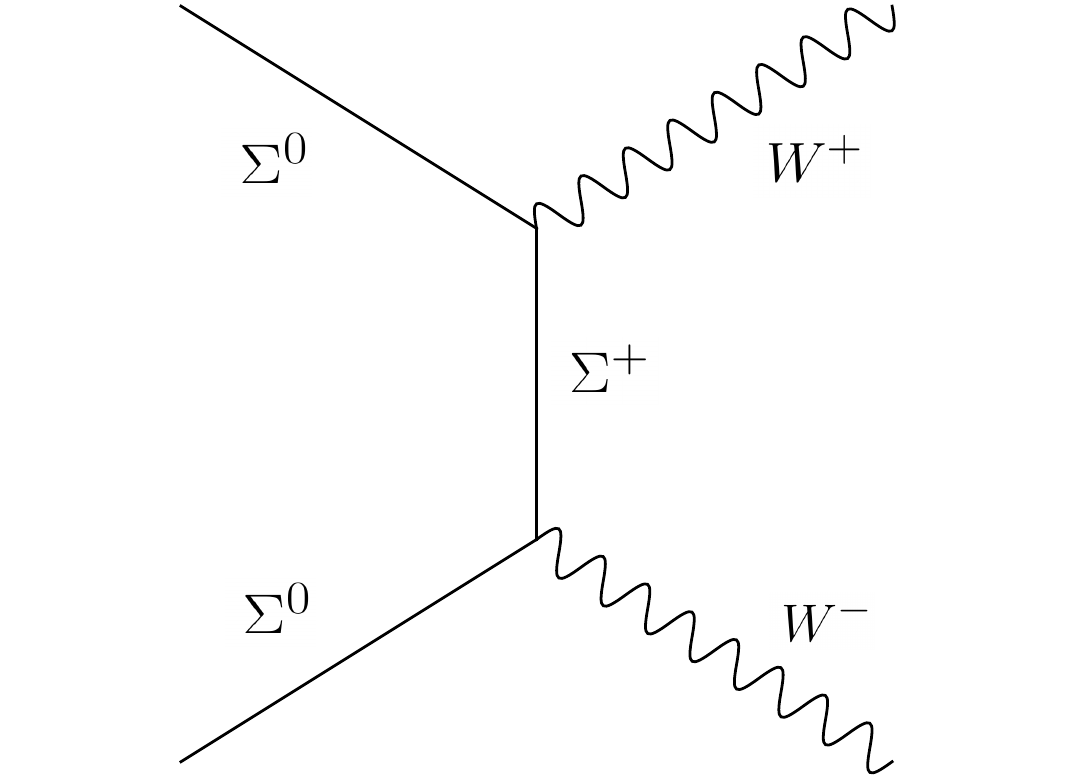}}

\subfigure[]{
		\centering
		\includegraphics[width = 0.35\textwidth]{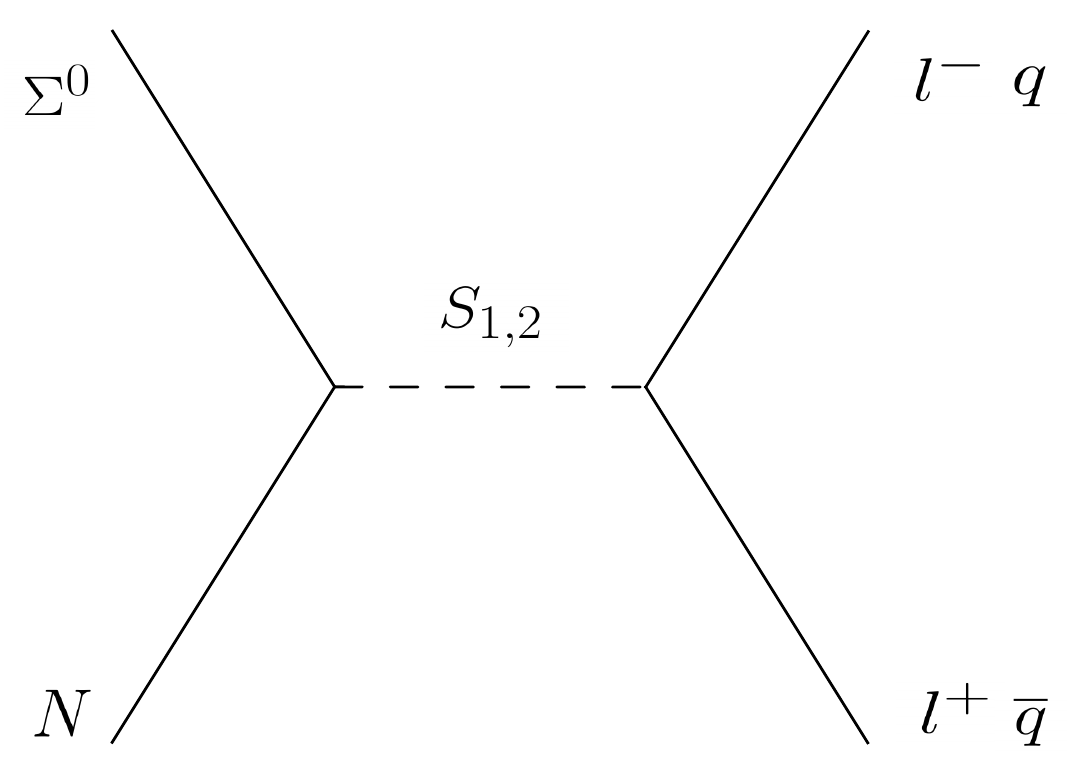}}
\subfigure[]{
		\centering
		\includegraphics[width = 0.35\textwidth]{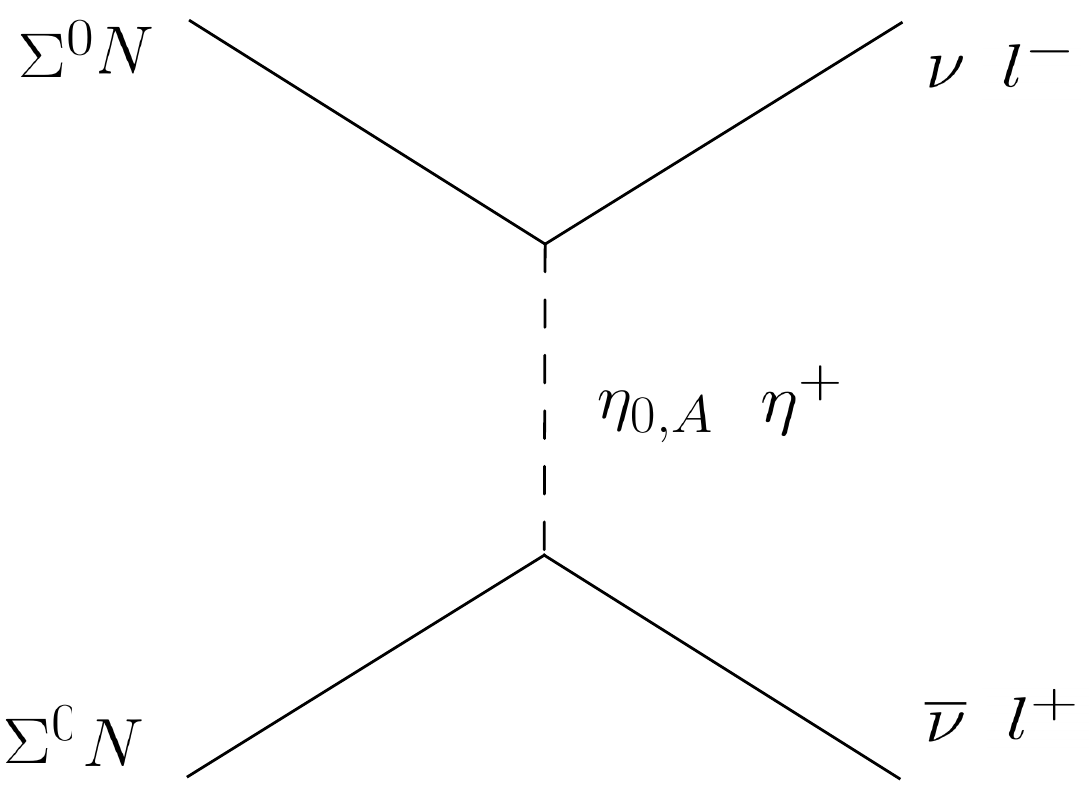}}

\subfigure[]{
		\centering
		\includegraphics[width = 0.35\textwidth]{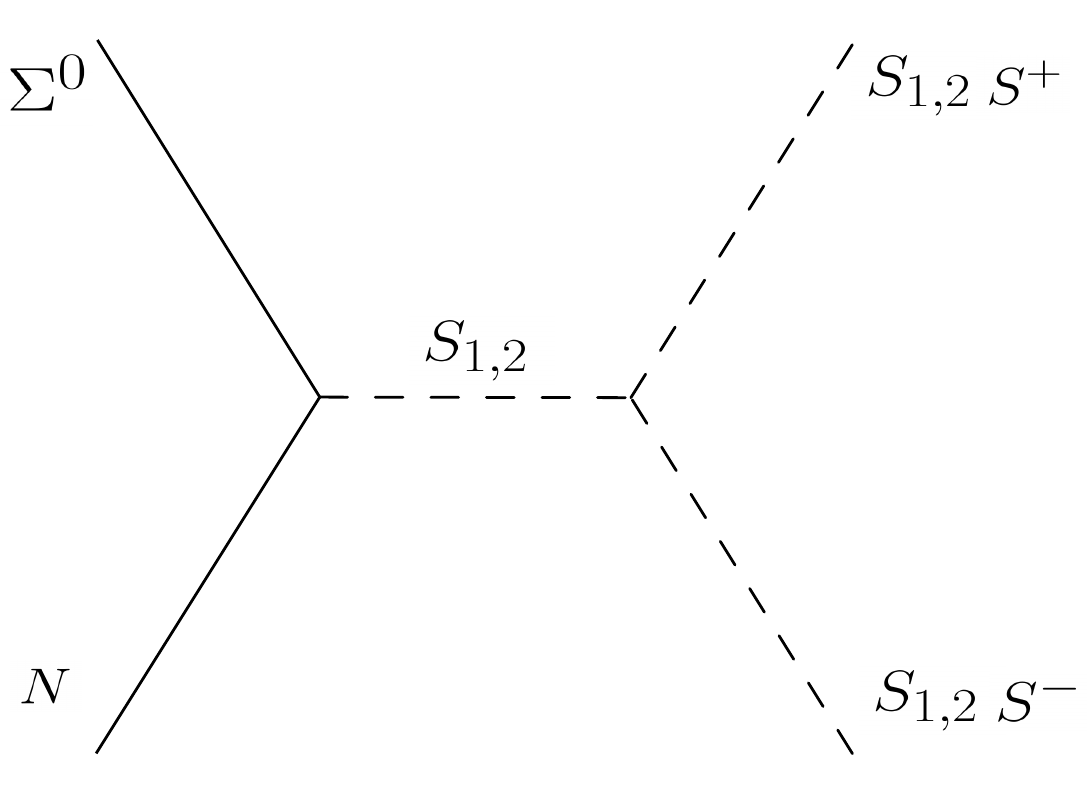}}
\subfigure[]{
		\centering
		\includegraphics[width = 0.35\textwidth]{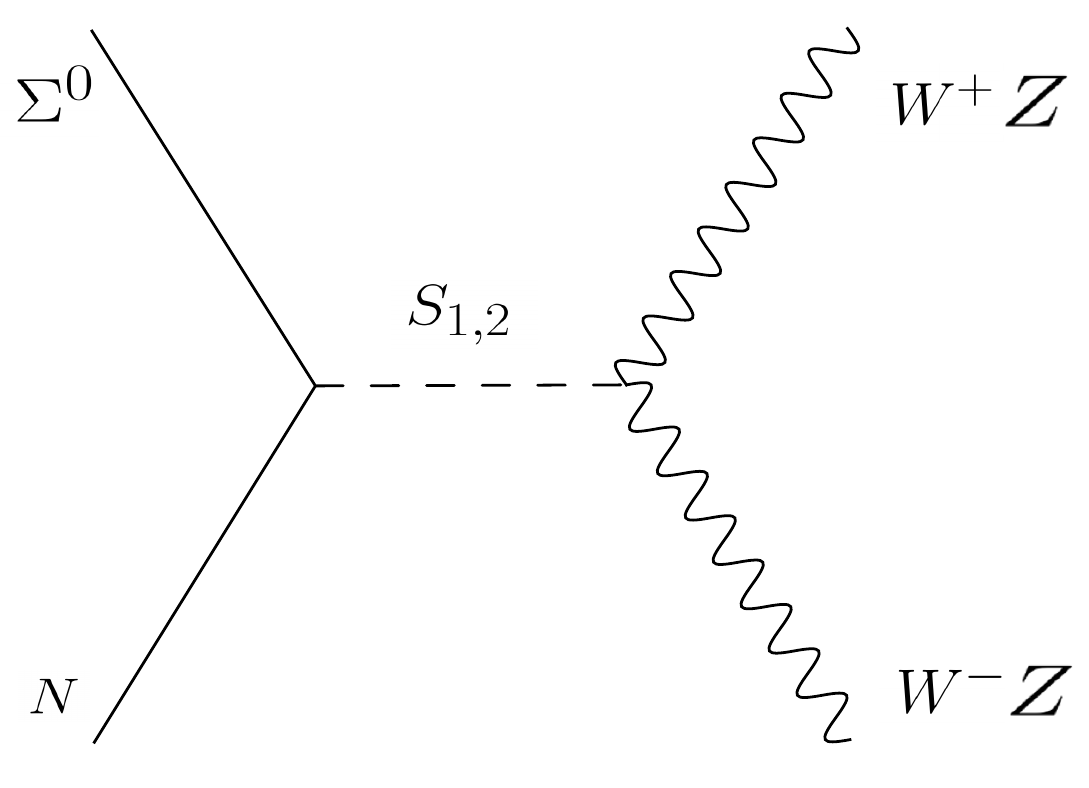}}

	\caption{$\Sigma^0$ and $N$ annihilation channels.\label{fig:ann}}
\end{figure}

As previously described the model contains two classes of potential
dark matter candidates.
One class are the $Z_2$ odd scalars: $\eta^0$ and $\eta^{A}$, when any
of them is the lightest $Z_2$ odd particle.
Their phenomenology is very close to the inert doublet dark matter
model~\cite{LopezHonorez:2006gr} or discrete dark matter
models~\cite{Hirsch:2010ru,Boucenna:2011tj}.
For this reason here we focus our analysis on the other candidates
which are the fermion states $\chi^0_i$.
In this case, the dark matter candidate is a mixed state between $N$
and $\Sigma^0$. This interplay brings an enriched dark matter
phenomenology with respect to models with only singlets or triplets.

\begin{figure}[tb]
\centering
\includegraphics[width=0.6\textwidth]{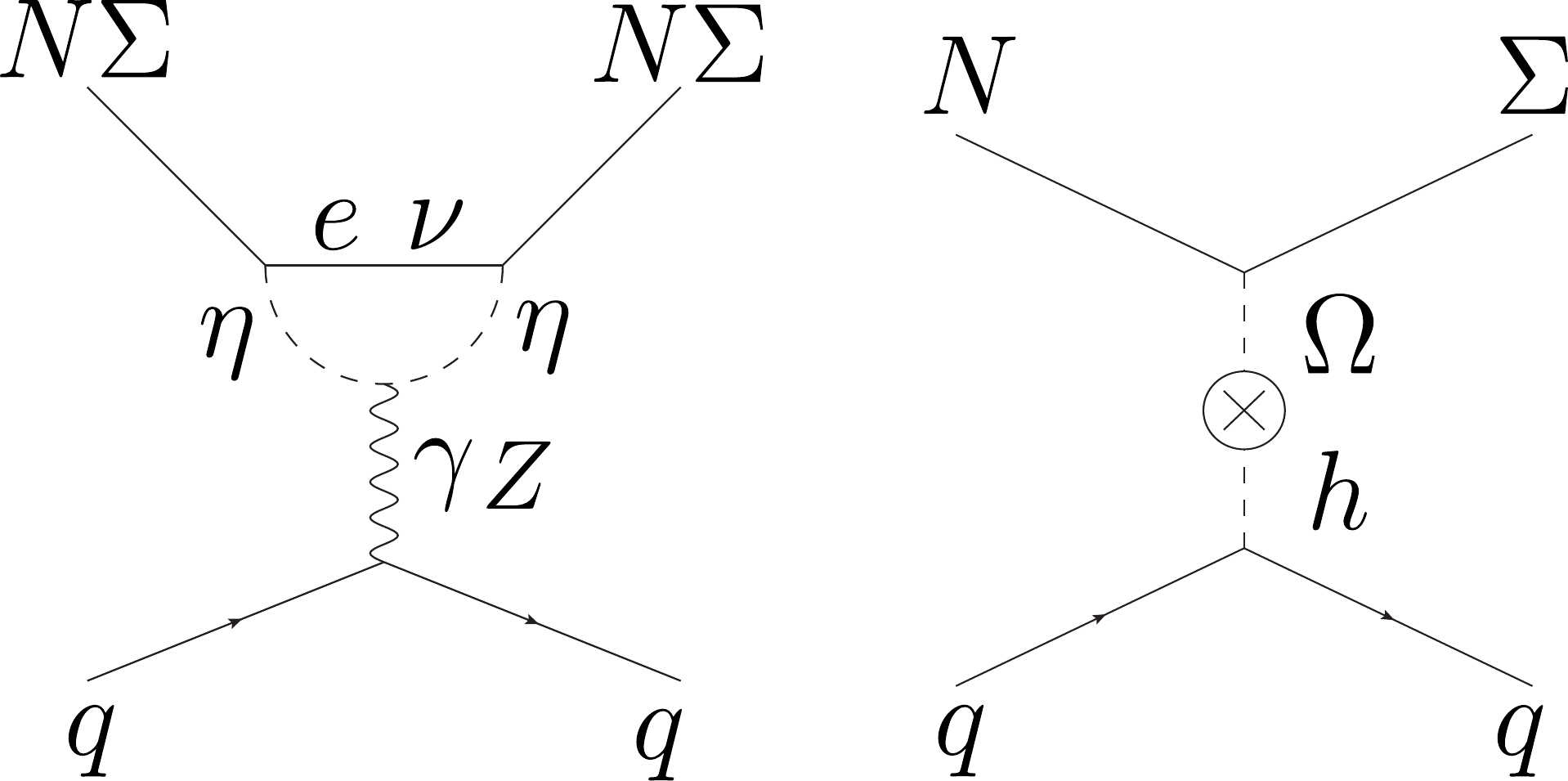}
\caption{Direct detection in pure triplet or pure singlet models (left
  panel) and in our mixed triplet-singlet case (right panel).}
\label{fig:dd0}

\end{figure}

For models with only fermion triplets as dark matter, equivalent in
our model to taking $M_N \to \infty$, the main constraints come from
the observed relic abundance (equation~\ref{eq:omegaDM}).
Coannihilations between $\Sigma^0$ and $\Sigma^{\pm}$ are efficient
processes due to the mass degeneracy between them, controlling the
relic abundance.
These processes force the dark matter mass to be 2.7~TeV. 
In addition, direct detection occurs only at the one loop
level~\cite{Cirelli:2005uq}, see Fig.~\ref{fig:dd0}.
Most of the corresponding features have been already studied
in~\cite{Ma:2008cu,Chao:2012sz}.
In figure~\ref{fig:coann}, we show the coannihilation channels present
in our model in terms of gauge eigenstates, except for the $Z_2$ even
scalars.
The dark matter mass can be much smaller for singlets fulfulling the
$\Omega_{\rm DM} h^2$ contraint.
However, processes related to direct detection are absent at tree
level~\cite{Schmidt:2012yg} for singlets too.
\begin{table}[tb]
\centering
\begin{tabular}{|c||c|}
\hline
\phantom{mmmmmm} Parameter \phantom{mmmmmm}& \phantom{mmmmmm}Range\phantom{mmmmmm}\\
\hline     
$M_N$ (GeV) & 1 -- $10^5$\\
$M_\Sigma$ (GeV) & 100 -- $10^5$\\
$m_{\eta^\pm}$ (GeV) & 100 -- $10^5$\\
$M_{\pm}$ (GeV) & 100 -- $10^4$\\
$|\lambda_i|$& $10^{-4}$  -- 1 \\
$|\lambda_i^{\eta,\Omega}|$& $10^{-4}$  -- 1 \\
$|Y_i|$ & $10^{-4}$ -- 1  \\
\hline     
\end{tabular}
\caption{Scanning parameter ranges. The remaing parameters are 
  calculated from this set.}
\label{tab:scan}
\end{table}

The presence of the scalar triplet $\Omega$ and its nonzero vev
induces a mixing between $\Sigma^0$ and $N$, implying coannihilations
that can be important when the dark matter has a large component of
$\Sigma^0$. This mixing also breaks the degeneracy between the mass
eigenstate fermions $\chi_{1}^0$ and $\chi^{\pm}$.
However, in this case, the mass degeneracy with the charged fermion
$\chi^{\pm}$ is increased and forces the dark matter to be
$\mathcal{O}({\rm TeV})$.
Other coannihilation processes occur when $M_N$ is also degenerate
with $M_{\Sigma}$.
For the opposite case, when $\chi^0$ is mainly $N$, the model
reproduces the phenomenology of the fermion singlet dark matter where
the main signature is the annihilation into neutrinos and charged
leptons (as in leptophilic dark matter) without any direct detection
prospective~\cite{Schmidt:2012yg}.
The potential scenarios present in the model have the best of
singlet-only or triplets-only scenarios and more.
In addition, the dark matter phenomenology includes new annihilation
and coannihilation channels when kinematically accessible.

The presence of the scalar triplet $\Omega$ also induces an
interaction between dark matter and quarks (direct detection) via the
exchange of neutral scalar $S_i (h^0, \Omega^{0})$, as illustrated in
In Fig~\ref{fig:ann}, we show the main diagrams of the model related
to indirect and direct searches.
The model can potentially produce the typical annihilation channels
appearing in generic weakly interactive massive particle dark matter
models.
Indeed, our dark matter candidate mimicks the Lightest Supersymmetric
Particle (neutralino) present in supergravity-like versions the
Minimal Supersymmetric Standard Model with R-parity conservation. The
latter would correspond here to our assumed $Z_2$ symmetry.

In order to study the dark matter phenomenology, we have implemented
the lagrangian (equation~\ref{eq:lagrangian}) using the standard codes
\mbox{{\tt LanHEP}}~\cite{lanhep:1996,lanhep:2009,lanhep:2010} and
{\tt Micromegas}~\cite{micromegas:2013}.
We scan the parameter space of the model within the ranges indicated
in Tab.~\ref{tab:scan}. We also take into account the following
constraints: 
perturbatibity and a Higgs--like
scalar at $\sim$~125~GeV.
Also we take into account the constraints from the relic
abundance~\cite{planck:2013} as well as the lower bound on the masses
of new non-colored charged particles coming from
LEP~\cite{L3:2001PhLB} and LHC~\cite{CMS:2012PhLB} collider searches,
roughly translated to $M_{\rm LEP} > 100 \, {\rm GeV}$.
We calculate the thermally averaged annihilation cross section
$\langle \sigma v \rangle$, and the spin independent cross section
$\sigma_{\rm SI}$.\\

\begin{figure}[tb]
	\centering
	\includegraphics[width = \textwidth]{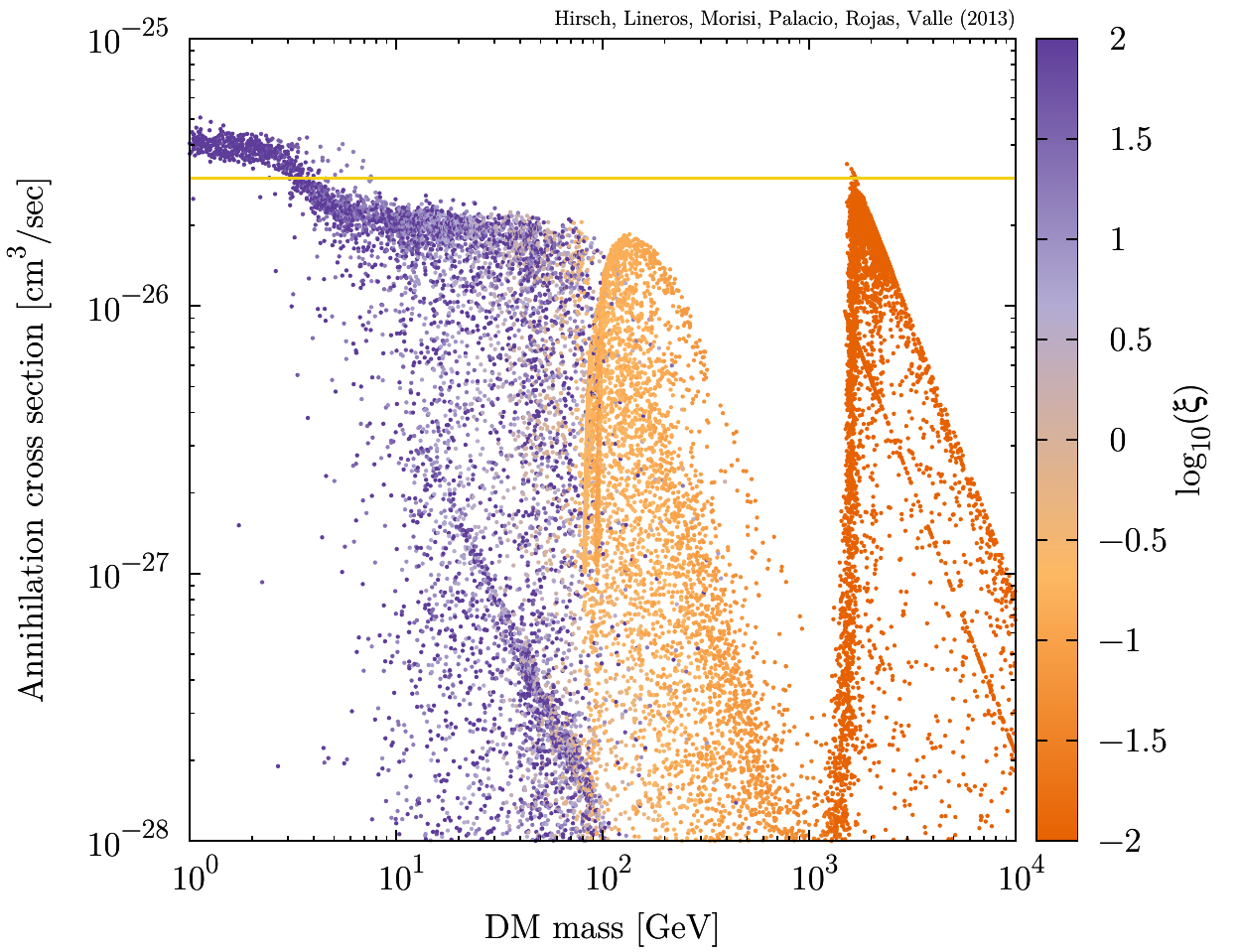}
	\caption{Annihilation cross section vs dark matter mass. Color
          scale represents $\log_{10}(\xi)$. Dark matter with masses
          larger than 1~TeV have a larger component of $\Sigma^0$,
          cases with masses lower than 20~GeV have larger component of
          $N$. The yellow line corresponds to the thermal value
          $3\times10^{-26} \textnormal{cm}^3/ \textnormal{sec}$.}
	\label{fig:anni_mass}
\end{figure}

\begin{figure}[tb]
	\centering
	\includegraphics[width = \textwidth]{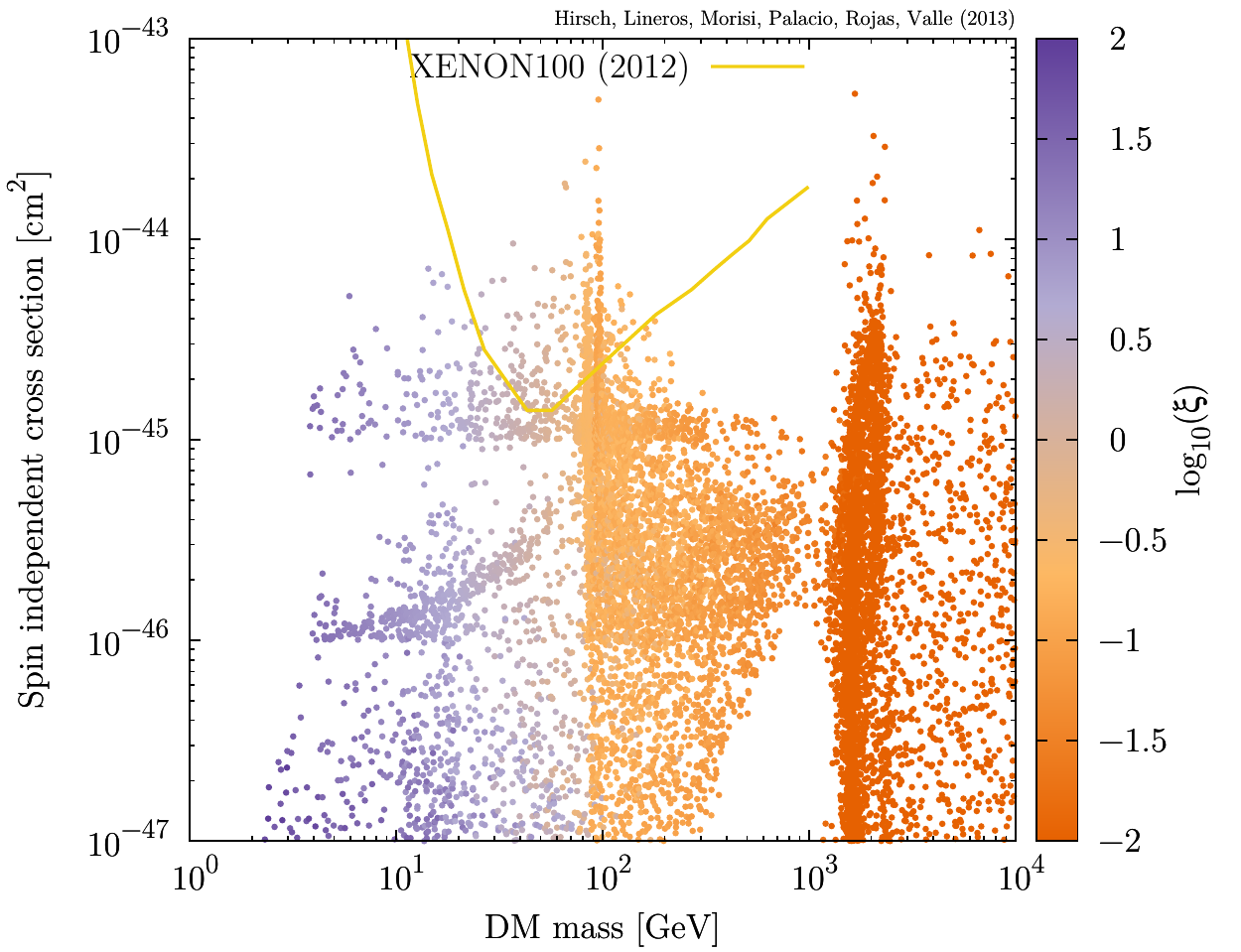}
	\caption{Spin independent cross section vs dark matter
          mass. Color scale is the same as in
          figure~\ref{fig:anni_mass}. The yellow line is the upper
          bound from XENON100 experiment~\cite{XENON:2012Ph}. }
	\label{fig:dd_mass}
\end{figure}

In figure~\ref{fig:anni_mass}, we present the results of the scan in
terms of the annihilation cross section versus the dark matter
mass. Moreover, we show in color scale the quantity:
\begin{equation}
	\xi = \frac{M_{\Sigma} - m_{\rm DM}}{m_{\rm DM}} \, ,
\end{equation} 
which estimates how degenerate is the dark matter mass with respect to
$M_{\Sigma}$. Small values of $\xi$ imply dark matter with a large
component of $\Sigma^0$ and large value implies a large component of
$N$. This quantity has implications for coannihilation processes
discussed previously.
We notice that regions with low dark matter masses ($< 20$~GeV) are
less degenerate mainly because $M_{\Sigma} > M_{\rm LEP}$.
In this region the dark matter contains a large component of $N$.
As expected, the TeV region is dominated by dark matter with large
component of $\Sigma^{0}$. The mass range 100--800 GeV is particularly
interesting because any of the new charged particles are accessible at
LHC.
Moreover, when the $\Sigma^0/N$ mixing is non-zero and $\displaystyle
m_{\rm DM} \simeq \frac{m_{S_i}}{2}$, the annihilation channels into
quarks and leptons are naturally enhanced due to the $s$-channel
resonance in the process:
\begin{equation}
  \chi_1^0 \chi_1^0 \to S_i \to f \bar{f} \, \to \langle \sigma v \rangle \propto \left(\frac{\sin(2 \, \alpha)}{(2 m_{\rm DM})^2 - m_{S_i}^2}\right)^2 \, .
\end{equation}
This is translated into higher expected fluxes of gamma--rays and
cosmic--rays for indirect searches as well as higher spin independent
cross section.\\


Now, turning to the direct detection perspectives, the plot of the
spin--independent cross section versus the dark matter mass is shown in
figure~\ref{fig:dd_mass}.
The scattering with quarks is described only with one diagram (the
exchange of scalars $S_i$), also shown in figure~\ref{fig:dd0}.
The size of the interaction will depend directly on the mixing
$\Sigma^0/N$.
For masses larger than 100 GeV, we observe an increase of
$\sigma_{SI}$ because maximal mixing can be obtained for $M_{N} \sim
M_{\Sigma}$ and for $Y_{\Omega} v_{\Omega} \ne 0$.
This does not occur for masses much lower to 100 GeV since the dark
matter becomes mainly a pure $N$.
Moreover, the model produces $\sigma_{\rm SI}$ large enough to be
observed in direct detection experiments such
XENON100~\cite{XENON:2012Ph} (yellow line).
%




Finally, we note that the new particles introduced in our model can be
kinematically accessible at the LHC. Here we briefly comment on
relevant production cross sections for the LHC.
Both, ATLAS~\cite{ATLAS-CONF-2013-019} and CMS~\cite{CMS:2012PhLB}
have searched for pair production of heavy triplet fermions:
$\Sigma^0+\Sigma^+$, deriving lower limits on $m_{\Sigma^{+}}$ of the
order of $m_{\Sigma^{+}} \gtrsim (180-210)$~GeV~\cite{CMS:2012PhLB}
and $m_{\Sigma^+} \gtrsim 245$~GeV~\cite{ATLAS-CONF-2013-019},
respectively. However, these bounds do not apply to our model, because
the final state topologies used in these searches, tri-leptons in case
of CMS~\cite{CMS:2012PhLB} and four charged leptons in ATLAS
\cite{ATLAS-CONF-2013-019}, are based on the assumption that
$\Sigma^0$ decays to the final states $\Sigma^0\to l^{\pm}l^{\mp}
+\nu/{\bar \nu}$.  As a result of the $Z_2$ symmetry present in our
model, however, the lightest fermion or scalar is stable and all
heavier $Z_2$-odd states will decay to this lightest state. Thus, the
intermediate states $\Sigma^{0} +\Sigma^{+}$ and $\Sigma^{-} +
\Sigma^{+}$, which have the largest production cross sections of all
new particles in our model, will not give rise to three and four
charged lepton signals.

Instead, the phenomenology of $\Sigma^0$ and $\Sigma^+$ depends on the
unknown mass ordering of fermions and scalars. Since we have assumed
in this paper that the lighter of the fermions is the dark matter, we
will discuss only this case here. Then, the phenomenology depends on
whether the lightest of the neutral fermions, $\chi^0_1$, is mostly
singlet or mostly triplet.
Consider first the case $\chi^0_1 \simeq \Sigma^{0}$. Then, from the
pair $\chi^0_1 + \Sigma^+$, only $\Sigma^+$ decays via $\Sigma^+ \to
\chi^0_1 + W^+$, where the $W^+$ can be on-shell or off-shell.
Thus, the final state consists mostly one charged lepton plus missing
energy.
The other possibility is pair production of $\Sigma^{+} + \Sigma^{-}$
via photon exchange, which leads to $l^++l^-$ plus missing energy.
In both cases, standard model backgrounds will be large and the LHC
data probably does not give any competitive limits yet.
We expect that LHC data at 14 TeV with increased statistics may
constrain part of the parameter space. A quantitative study would
require a MonteCarlo analisys which is beyond the scope of this work.

Conversely, for the case $\chi^0_2 \simeq \Sigma^{0}$, the $\chi^0_2$
will decay to $\chi^0_1$ plus either one on-shell or off-shell Higgs
state, depending on kinematics. In this case the final state will be
one charged lepton plus up to four b-jets plus missing momentum. This
topology is not covered by any searches at the LHC so far, as far as
we are aware.

Also, the new neutral and charged scalars can be searched for at the
LHC. All possible signals have, however, rather small production cross
sections. Neither $\eta$ nor $\Omega$ have couplings to quarks and
only $\Omega$ (both charged and neutral) can be produced at the LHC
due to its mixing with the \sm Higgs field $\phi$. Final states will
be very much SM-Higgs like, but the event numbers will depend
quadratically on this mixing, which supposedly is a small number,
since the observed state with a mass of roughly $(125-126)$ GeV
behaves rather closely like A \sm Higgs. Searches for a heavier state
with \sm like Higgs properties~\cite{Chatrchyan:2013yoa} exclude
scalars with standard coupling strength now up to roughly 700
GeV. However, upper limits on $\sin^2(\theta)$ in the mass range
$(130-700)$ GeV are currently only of the order $(0.2-1.0)$. The next
run at the LHC, with its projected luminosity of order ${\cal L}
\simeq (100-300)$ fb$^{-1}$, should allow to probe much smaller mixing
angles.

\section{Conclusions}
\label{sec:concl}

We have presented a next-to minimal extension of the \sm including new
$Z_2$-odd majorana fermions, one singlet $N$ and one triplet $\Sigma$
under weak SU(2), as well as a $Z_2$-odd scalar doublet $\eta$. We
also include a $Z_2$-even triplet scalar $\Omega$ in order induce the
mixing in the fermionic sector $N$--$\Sigma$.
The solar and atmospheric neutrino mass scales are then generated at
one-loop level, with the lightest neutrino remaining massless.
%
%
This way our model combines the ingredients present in
Refs.~\cite{Ma:2006km,Ma:2008cu} with a richer phenomenology.

The unbroken $Z_2$ symmetry implies that the lightest $Z_2$-odd
particle is stable and may play the role of dark matter. We analyze
the viability of the model using state-of-art codes for dark matter
phenomenology.
We focus our attention to the fermionic dark matter case. 
The mixing between $N$ and the neutral component of $\Sigma$ relaxes
the effects of coannihilations between the dark matter candidate and
the charged component of $\Sigma$.
In the pure triplet case, the dark matter mass is forced to be 2.7~TeV
in order to reproduce the observed dark matter abundance value.
However, in the presence of mixing the effect of coannihilations is
weaker, allowing for a reduced dark matter mass down to the GeV range.
Thanks to that, the charged $\Sigma$ can be much lighter than in the
pure triplet case, openning the possibility of new signatures at
colliders such as the LHC. In addition, the dark matter candidate can
interact with quarks at tree level and then produce direct detection
signal that may be observed or constrained in current direct searches
experiments such XENON100.

\section*{Acknowledgments}

This work was supported by the Spanish MINECO under grants
FPA2011-22975 and MULTIDARK CSD2009-00064 (Consolider-Ingenio 2010
Programme), by Prometeo/2009/091 (Generalitat Valenciana), and by the
EU ITN UNILHC PITN-GA-2009-237920.
S.M. thanks to DFG grant WI 2639/4-1 for financial support.
N.R. thanks to CONICYT doctoral grant, Marco A. D\'{\i}az for useful
discussions and comments, the EPLANET grant for funding the stay in
Valencia, and the IFIC--AHEP group in Valencia for the hospitality.
R.L. also thanks to V.~\c{T}\u{a}ranu for her support.

\appendix
\section{Appendix}
\subsection{Approximations for Neutrino Masses.} 
\label{sec:approx}

Starting from the equation~\ref{eq:numass}, one can perform some approximations
to examine neutrino masses for cases of interest, for example, cases with one of 
the $\chi^0_1$ masses being the lightest between $\chi^0_2$, 
$\eta_{0,A}$, $\Sigma_{\pm}$ and $\Omega_{0,\pm}$.\\

One wants to establish the relation between neutrino masses and the other
parameters in the lagrangian in a suitable form. In principle, neutrino 
masses depend on the masses of neutral $\eta$ fields and the masses of 
the $\chi^0$, but the dependence of the parameters of the scalar 
sector is more complicated, given the structure of the masses of 
the $\eta$ fields (see equations~\ref{eta0} and \ref{etaA}). One can
take these equations and write them in the following way:
\begin{eqnarray} 
m_{\eta_0}^2 &=& m_0^2 + \lambda_5 v_h^2 \, ,\\
m_{\eta_A}^2 &=& m_0^2 - \lambda_5 v_h^2 \, .
\end{eqnarray}

Where $m_0^2$ is a complicated function of the parameters of the scalar 
potential. One can write the equation~\ref{eq:Iformula} as follows:
\begin{eqnarray}
I_k &=& - M_k \left( \frac{m_0^2 + \lambda_5 v_h^2}{M_k^2 - m_0^2 - \lambda_5 v_h^2} \right) \log \left( \frac{m_0^2 + \lambda_5 v_h^2}{M_k^2} \right) \nonumber \\
 && + M_k \left( \frac{m_0^2 - \lambda_5 v_h^2}{M_k^2 - m_0^2 + \lambda_5 v_h^2} \right) \log \left( \frac{m_0^2 - \lambda_5 v_h^2}{M_k^2} \right)\, .  
\end{eqnarray}

{}One can identify two interesting limit cases. 
When $\lambda_5 v_h^2 \ll M_k^2 \approx m_0^2$ then the $I_k$ function can be written as:
\begin{eqnarray}
I_k &=& \frac{2\lambda_5 v_h^2}{M_k} \, .
\end{eqnarray}
Therefore, the neutrino mass matrix in this approximation is given by:
\begin{eqnarray}
M_{\alpha\beta}^\nu &=& \sum_{\sigma = 1,2} \frac{h_{\alpha \sigma}h_{\beta \sigma}}{8\pi^2} \frac{\lambda_5 v_h^2}{M_\sigma} \, . \label{eq:numass2}
\end{eqnarray}

The other case is given by  $\lambda_5 v_h^2\, ,\, M_k^2 \ll m_0^2$, the
procedure is not difficult, the result is:
\begin{eqnarray}
I_k &=& \frac{2\lambda_5 v_h^2}{m_0^2}M_k\, .
\end{eqnarray}
In this case, the neutrino mass matrix is given by:
\begin{eqnarray}
M_{\alpha\beta}^\nu &=& \sum_{\sigma = 1,2} \frac{h_{\alpha \sigma}h_{\beta \sigma}}{8\pi^2} \frac{\lambda_5 v_h^2}{m_0^2}M_k \, . \label{eq:numass3}
\end{eqnarray}

\subsection{Minimization conditions}
\label{app:tad}

{}The tadpole equations were computed in order to find the minimum of
the scalar potential, thus, the linear terms of the scalar potential
at tree level can be written as:

\begin{eqnarray}
V_{(1)} &=& t_h h_0 + t_\eta \eta_0 + t_\Omega \Omega_0
\end{eqnarray}

{}Where the tadpoles are:
\begin{eqnarray}
t_h &=& v_h\left( -m_1^2 + \frac{1}{2}\lambda_1 v_h^2 + \frac{1}{2}\left( \lambda_3 + \lambda_4 + \lambda_5 \right)v_\eta^2 \right) \\
t_\eta &=& v_\eta\left( m_2^2 + \frac{1}{2}\lambda_2 v_\eta^2 + \frac{1}{2}\left( \lambda_3 + \lambda_4 + \lambda_5 \right)v_h^2 \right) \label{eq:tadeta}\\
t_\Omega &=& -M_\Omega^2 v_\Omega - \mu_1 v_h^2 + \left(2\lambda_1^\Omega + \lambda_4^\Omega \right)v_h^2 v_\Omega + \nonumber \\
          & & 8\left( 2\lambda_2^\Omega + \lambda_3^\Omega \right)v_h^2 v_\Omega^3 + \mu_2 v_\eta^2 + \left( 2\lambda_1^\eta + \lambda_4^\eta \right)v_\Omega^2 v_\eta^2
\end{eqnarray}

{}In order to have an $Z_2$ invariant vacuum, the vev $v_\eta$ has
to vanish, which is extracted from the equation \ref{eq:tadeta}. For
the vev $v_h$, one can choose the value to be nonzero solving the equation
in the parenthesis, in equal manner, one obtain the vev $v_\Omega$, in 
terms of the other parameters of the potential. \\

{}The numerical values of the vevs $v_h$ and $v_\Omega$ are restricted to
reproduce the measured values of gauge boson masses, this allows to have the
value for $v_h \sim 246$~GeV, and $v_\Omega \leq 7$~GeV, as one can see in
the section \ref{sec:scalars}.

\newcommand{\apjl}{Astrophys. J. Lett.}
\newcommand{\apjs}{Astrophys. J. Suppl. Ser.}
\newcommand{\aap}{Astron. \& Astrophys.}
\newcommand{\aj}{Astron. J.}
\newcommand{\araa}{Ann. Rev. Astron. Astrophys. } 
\newcommand{\mnras}{Mon. Not. R. Astron. Soc.}
\newcommand{\physrep}{Phys. Rept.}
\newcommand{\jcap}{JCAP}
\newcommand{\prl}{PRL}
\newcommand{\prd}{PRD}

\bibliographystyle{JHEP}

\bibliography{refs.bib}

\providecommand{\href}[2]{#2}\begingroup\raggedright\begin{thebibliography}{10}

\bibitem{Schwetz:2011zk}
T.~Schwetz, M.~Tortola, and J.~Valle, {\it {Where we are on $\theta_{13}$:
  addendum to `Global neutrino data and recent reactor fluxes: status of
  three-flavour oscillation parameters'}},  {\em New J.Phys.} {\bf 13} (2011)
  109401, [\href{http://xxx.lanl.gov/abs/1108.1376}{{\tt arXiv:1108.1376}}].

\bibitem{Tortola:2012te}
D.~Forero, M.~Tortola, and J.~Valle, {\it {Global status of neutrino
  oscillation parameters after Neutrino-2012}},  {\em Phys.Rev.} {\bf D86}
  (2012) 073012, [\href{http://xxx.lanl.gov/abs/1205.4018}{{\tt
  arXiv:1205.4018}}].

\bibitem{Schechter:1981cv}
J.~Schechter and J.~Valle, {\it {Neutrino Decay and Spontaneous Violation of
  Lepton Number}},  {\em Phys.Rev.} {\bf D25} (1982) 774.

\bibitem{Foot:1988aq}
R.~Foot, H.~Lew, X.~He, and G.~C. Joshi, {\it {Seesaw neutrino masses induced
  by a triplet of leptons}},  {\em Z.Phys.} {\bf C44} (1989) 441.

\bibitem{Ma:2006km}
E.~Ma, {\it {Verifiable radiative seesaw mechanism of neutrino mass and dark
  matter}},  {\em Phys.Rev.} {\bf D73} (2006) 077301,
  [\href{http://xxx.lanl.gov/abs/hep-ph/0601225}{{\tt hep-ph/0601225}}].

\bibitem{Ma:2008cu}
E.~Ma and D.~Suematsu, {\it {Fermion Triplet Dark Matter and Radiative Neutrino
  Mass}},  {\em Mod.Phys.Lett.} {\bf A24} (2009) 583--589,
  [\href{http://xxx.lanl.gov/abs/0809.0942}{{\tt arXiv:0809.0942}}].

\bibitem{Ellis:1988zy}
J.~R. Ellis and F.~Pauss, {\it {SEARCHES FOR NEW PHYSICS}},  {\em
  Adv.Ser.Direct.High Energy Phys.} {\bf 4} (1989) 269--322.

\bibitem{L3:2001PhLB}
{L3 Collaboration}, {\it {Search for heavy neutral and charged leptons in
  e$^{+}$e$^{-}$ annihilation at LEP}},  {\em Physics Letters B} {\bf 517}
  (Sept., 2001) 75--85, [\href{http://xxx.lanl.gov/abs/hep-ex/01}{{\tt
  hep-ex/01}}].

\bibitem{CMS:2012PhLB}
{CMS Collaboration}, {\it {Search for heavy lepton partners of neutrinos in
  proton-proton collisions in the context of the type III seesaw mechanism}},
  {\em Physics Letters B} {\bf 718} (Dec., 2012) 348--368,
  [\href{http://xxx.lanl.gov/abs/1210.1797}{{\tt arXiv:1210.1797}}].

\bibitem{planck:2013}
{Planck Collaboration}, {\it {Planck 2013 results. XVI. Cosmological
  parameters}},  {\em ArXiv e-prints} (Mar., 2013)
  [\href{http://xxx.lanl.gov/abs/1303.5076}{{\tt arXiv:1303.5076}}].

\bibitem{Kubo:2006yx}
J.~Kubo, E.~Ma, and D.~Suematsu, {\it {Cold Dark Matter, Radiative Neutrino
  Mass, mu $\rightarrow$ e gamma, and Neutrinoless Double Beta Decay}},  {\em
  Phys.Lett.} {\bf B642} (2006) 18--23,
  [\href{http://xxx.lanl.gov/abs/hep-ph/0604114}{{\tt hep-ph/0604114}}].

\bibitem{Gunion:1989ci}
J.~Gunion, R.~Vega, and J.~Wudka, {\it {Higgs triplets in the standard model}},
   {\em Phys.Rev.} {\bf D42} (1990) 1673--1691.

\bibitem{Gunion:1989we}
J.~F. Gunion, H.~E. Haber, G.~L. Kane, and S.~Dawson, {\it {THE HIGGS HUNTER'S
  GUIDE}},  {\em Front.Phys.} {\bf 80} (2000) 1--448.

\bibitem{Passarino:1978jh}
G.~Passarino and M.~Veltman, {\it {One Loop Corrections for e+ e- Annihilation
  Into mu+ mu- in the Weinberg Model}},  {\em Nucl.Phys.} {\bf B160} (1979)
  151.

\bibitem{Schechter:1980gr}
J.~Schechter and J.~W.~F. Valle, {\it Neutrino masses in su(2) x u(1)
  theories},  {\em Phys. Rev.} {\bf D22} (1980) 2227.

\bibitem{Casas:2001sr}
J.~Casas and A.~Ibarra, {\it {Oscillating neutrinos and muon $\rightarrow$ e,
  gamma}},  {\em Nucl.Phys.} {\bf B618} (2001) 171--204,
  [\href{http://xxx.lanl.gov/abs/hep-ph/0103065}{{\tt hep-ph/0103065}}].

\bibitem{LopezHonorez:2006gr}
L.~Lopez~Honorez, E.~Nezri, J.~F. Oliver, and M.~H. Tytgat, {\it {The Inert
  Doublet Model: An Archetype for Dark Matter}},  {\em JCAP} {\bf 0702} (2007)
  028, [\href{http://xxx.lanl.gov/abs/hep-ph/0612275}{{\tt hep-ph/0612275}}].

\bibitem{Hirsch:2010ru}
M.~Hirsch, S.~Morisi, E.~Peinado, and J.~Valle, {\it {Discrete dark matter}},
  {\em Phys.Rev.} {\bf D82} (2010) 116003,
  [\href{http://xxx.lanl.gov/abs/1007.0871}{{\tt arXiv:1007.0871}}].

\bibitem{Boucenna:2011tj}
M.~Boucenna, M.~Hirsch, S.~Morisi, E.~Peinado, M.~Taoso, et~al., {\it
  {Phenomenology of Dark Matter from $A_4$ Flavor Symmetry}},  {\em JHEP} {\bf
  1105} (2011) 037, [\href{http://xxx.lanl.gov/abs/1101.2874}{{\tt
  arXiv:1101.2874}}].

\bibitem{Cirelli:2005uq}
M.~Cirelli, N.~Fornengo, and A.~Strumia, {\it {Minimal dark matter}},  {\em
  Nucl.Phys.} {\bf B753} (2006) 178--194,
  [\href{http://xxx.lanl.gov/abs/hep-ph/0512090}{{\tt hep-ph/0512090}}].

\bibitem{Chao:2012sz}
W.~Chao, {\it {Dark Matter, LFV and Neutrino Magnetic Moment in the Radiative
  Seesaw Model with Triplet Fermion}},
  \href{http://xxx.lanl.gov/abs/1202.6394}{{\tt arXiv:1202.6394}}.

\bibitem{Schmidt:2012yg}
D.~Schmidt, T.~Schwetz, and T.~Toma, {\it {Direct Detection of Leptophilic Dark
  Matter in a Model with Radiative Neutrino Masses}},  {\em Phys.Rev.} {\bf
  D85} (2012) 073009, [\href{http://xxx.lanl.gov/abs/1201.0906}{{\tt
  arXiv:1201.0906}}].

\bibitem{lanhep:1996}
A.~V. {Semenov}, {\it {LanHEP - a package for automatic generation of Feynman
  rules in gauge models}},  {\em ArXiv High Energy Physics - Phenomenology
  e-prints} (Aug., 1996) [\href{http://xxx.lanl.gov/abs/hep-ph/96}{{\tt
  hep-ph/96}}].

\bibitem{lanhep:2009}
A.~V. {Semenov}, {\it {LanHEP a package for the automatic generation of Feynman
  rules in field theory. Version 3.0}},  {\em Computer Physics Communications}
  {\bf 180} (Mar., 2009) 431--454,
  [\href{http://xxx.lanl.gov/abs/0805.0555}{{\tt arXiv:0805.0555}}].

\bibitem{lanhep:2010}
A.~{Semenov}, {\it {LanHEP - a package for automatic generation of Feynman
  rules from the Lagrangian. Updated version 3.1}},  {\em ArXiv e-prints} (May,
  2010) [\href{http://xxx.lanl.gov/abs/1005.1909}{{\tt arXiv:1005.1909}}].

\bibitem{micromegas:2013}
G.~{Belanger}, F.~{Boudjema}, A.~{Pukhov}, and A.~{Semenov}, {\it
  {micrOMEGAs3.1 : a program for calculating dark matter observables}},  {\em
  ArXiv e-prints} (May, 2013) [\href{http://xxx.lanl.gov/abs/1305.0237}{{\tt
  arXiv:1305.0237}}].

\bibitem{XENON:2012Ph}
E.~{Aprile} et~al., {\it {Dark Matter Results from 225 Live Days of XENON100
  Data}},  {\em Physical Review Letters} {\bf 109} (Nov., 2012) 181301,
  [\href{http://xxx.lanl.gov/abs/1207.5988}{{\tt arXiv:1207.5988}}].

\bibitem{ATLAS-CONF-2013-019}
{\it Search for type-iii seesaw model heavy fermions in events with four
  charged leptons using 5.8/fb of sqrt(s)=8 tev data with the atlas detector},
  Tech. Rep. ATLAS-CONF-2013-019, CERN, Geneva, Sep, 2013.

\bibitem{Chatrchyan:2013yoa}
{\bf CMS Collaboration} Collaboration, S.~Chatrchyan et~al., {\it {Search for a
  standard-model-like Higgs boson with a mass in the range 145 to 1000 GeV at
  the LHC}},  {\em Eur.Phys.J.} {\bf C73} (2013) 2469,
  [\href{http://xxx.lanl.gov/abs/1304.0213}{{\tt arXiv:1304.0213}}].

\end{thebibliography}\endgroup

\end{document}